\pdfoutput=1

\documentclass[aps,prb,reprint,showpacs,superscriptaddress]{revtex4-1}

\usepackage{graphicx}
\usepackage{amsmath}
\usepackage{amssymb}
\usepackage{dcolumn}
\usepackage{dsfont}
\usepackage{latexsym}
\usepackage{rotating}
\usepackage{color}
\usepackage{latexsym}
\usepackage{bbm}
\usepackage[caption=false]{subfig}
\usepackage{float}
\usepackage{epsfig}
\usepackage{psfrag}
\usepackage{natbib}
\usepackage{bm}
\usepackage{amsthm}
\usepackage{eucal}
\usepackage{mathrsfs}
\usepackage{url}

\usepackage{color} 

\usepackage{hyperref}
\hypersetup{
colorlinks=true,final=true,
        linkcolor=blue,
        citecolor=blue,
        filecolor=blue,
        urlcolor=blue,
}

\begin{document}

\title{Influence of structural disorder and Coulomb interactions in the superconductor-insulator transition applied to boron doped diamond}

\author{R. McIntosh}
\affiliation{Nano-scale Transport Physics Laboratory, School of Physics, and Centre of Excellence in Strong Materials, 
\\University of the
Witwatersrand, Private Bag 3, WITS 2050, Johannesburg, South Africa}
\author{N. Mohanta}
\affiliation{Center for Electronic Correlations and Magnetism, Theoretical Physics III,
Institute of Physics, University of Augsburg, 86135 Augsburg, Germany}
\author{A. Taraphder}
\affiliation{Department of Physics and Centre for Theoretical Studies, Indian Institute of Technology, 
\\Kharagpur 721302, India}
\author{S. Bhattacharyya}
\email[]{Somnath.Bhattacharyya@wits.ac.za} 
\affiliation{Nano-scale Transport Physics Laboratory, School of Physics, and Centre of Excellence in Strong Materials, 
\\University of the
Witwatersrand, Private Bag 3, WITS 2050, Johannesburg, South Africa}

\begin{abstract}

The influence of disorder, both structural (non-diagonal) and on-site (diagonal), is studied through the inhomogeneous Bogoliubov-de Gennes (BdG) theory in narrow-band disordered superconductors with a view towards understanding superconductivity in boron doped diamond (BDD) and boron-doped nanocrystalline diamond (BNCD) films. We employ the attractive Hubbard model within the mean field approximation, including the Coulomb interaction between holes in the narrow acceptor band.  We study substitutional boron incorporation in a triangular lattice, with disorder in the form of random potential fluctuations at the boron sites. The role of structural disorder was studied through non-uniform variation of the tight-binding coupling parameter where, following experimental findings, we incorporate the concurrent increase in structural disorder with increasing boron concentration. We illustrate stark differences between the effects of structural and on-site disorder and show that structural disorder has a much greater effect on the density of states, mean pairing amplitude and superfluid density than on-site potential disorder. We show that structural disorder can increase the mean pairing amplitude while the spectral gap in the density of states decreases with states eventually appearing within the spectral gap for high levels of disorder. This study illustrates how the effects of structural disorder can explain some of the features found in superconducting BDD and BNCD films such as a tendency towards saturation of the $T_{c}$ with boron doping and deviations from the expected BCS theory in the temperature dependence of the pairing amplitude and spectral gap.

\end{abstract}

\maketitle



\section{\label{intro}Introduction}

\par{The interplay of superconductivity and disorder close to the Anderson-Mott transition has long been of significant fundamental interest in understanding quantum phase transitions and emergent phenomena \cite{AMTrans,AndTrans,GantRev}. The pivotal role of disorder in these phase changes has come to the forefront as the influence of different forms of disorder are widely studied \cite{Denteneer2001,Jiang2013,Trivedi1996R,SacepePRL2008,Potirniche2014,Mohanta2015}. Studies of 2$D$ localized superconductors using the Hubbard model with Anderson on-site disorder have shown that the spectral gap remains finite even at very high levels of disorder \cite{Ghosal1998, Chatterjee2008582,Ghosal2001}. The transition from macroscopic superconducting coherence to localized regions with a finite pairing amplitude interspersed by insulating regions has been studied using the BdG theory as well as Monte Carlo analysis \cite{Ghosal2001}. However, the Coulomb interaction, which is prevalent in many systems close to the Anderson-Mott transition is often neglected \cite{Trivedi1996R}. While the role of structural disorder has been examined in the context of localization of electrons, \cite{Cerovski1999} many aspects of the role of structural disorder in the superconductor-insulator transition are yet to be studied.}

\par{As an unconventional, disordered covalent superconductor boron doped diamond is of fundamental interest in studying the superconductor-insulator transition. Boron doped diamond is a type II superconductor, with a $ T_{c} $ as high as $ 10 $ $K$ \cite{Takano2004,PhysRevLett.96.097006}. Immediately following the unexpected discovery of superconductivity in BDD came along a proposal that BDD may be a new form of high $T_{c}$ superconductor, based expectedly on the resonating valence band (RVB) pairing mechanism \cite{Baskaran}. However, experimental evidence so far strongly suggests a phonon mediated pairing \cite{Achatz2010814,Dubrovinskaia,PhysRevLett.96.097006} without any evidence for the involvement of spins.} 

\par{Diamond is intrinsically a high band-gap insulator with a band-gap of $ 5.5$ $eV $. Boron can be  substitutionally incorporated in diamond, creating a deep, narrow acceptor band $ 0.37 $ $eV$ above the valence band \cite{Ekimov2004, PhysRevLett.93.237004}. While the acceptor states are three fold degenerate, the small spin-orbit coupling ($ 6 $ $meV$) as well as the random distribution of boron impurities is thought to lift this degeneracy \cite{Baskaran}. The transition with boron doping from insulator to superconductor has been studied in detail \cite{Kawano2010}, with the critical concentration found to be around $ 3 \times 10^{20} $ $cm^{-3}$. To explain the transition with boron doping, density functional theory studies have indicated a rigid band shift of the Fermi level with doping, however experimental studies suggest impurity states without a rigid band \cite{Shirakawa}. It has been illustrated that the $ T_{c} $ in boron doped diamond is limited by the low density of states at the Fermi energy \cite{PhysRevLett.93.237004}.}

\par{Photoemission studies of polycrystalline boron doped diamond have revealed a superconducting gap in the range expected for a weakly coupled Bardeen-Cooper-Schrieffer (BCS) superconductor while the characteristic broadening of the spectra is attributed to disorder \cite{Okazaki2011582}. Scanning transmission spectroscopy studies of BNCD have indicated regions which exhibit superconductivity consistent with the weak BCS limit interspersed with regions which are non-BCS like \cite{AchatzPSS}. In addition, BNCD grains have also shown broadening of the gap, believed to be related to disorder,  as well as to the distribution of grains with different  onset of $T_{c}$ \cite{PhysRevB.82.033306}. The sharp differential conductance peaks (referred to as coherence peaks) associated with s-wave superconductivity have also been observed \cite{PhysRevLett.96.097006}. Temperature dependent studies of the gap showed that the differential conductance conformed in some regions with BCS theory for weak coupling while the temperature dependence did not follow the conventional BCS dependence in other regions \cite{AchatzPSS}.

Theoretical studies employing the coherent potential approximation, with disorder in the form of a random potential at the boron sites, predict an exponentially increasing $ T_{c} $ with boron concentration \cite{Ohta2007121, Shirakawa} for disordered boron-doped crystalline diamond while experimental studies show an initially sharp increase in the $T_{c}$ with boron incorporation followed by a less prominent increase or even saturation \cite{Kawano2010,Klein2007,BustarretPRL,Achatz2010814} indicating deviations from what would be expected from the standard BCS dependence \cite{Winzer200565}. The Berlitz theory however has been shown to result in saturation of the $T_{c}$ which is in good agreement with experimental findings \cite{Bretislav2009}. These studies have indicated that the Coulomb interaction between holes in the acceptor band should be significant (although it has not been included in the calculations) \cite{Baskaran,Ohta2007121, Shirakawa}. Density functional theory studies assuming phonon mediated superconductivity with random substitutional boron have suggested that the $T_{c}$ for BDD could be as high as 55 $ K $ \cite{PhysRevB.77.064518}, providing motivation for understanding the role disorder plays in limiting the $T_{c}$. In addition, $ab-initio$ calculations have shown a difference in the bond-length associated with $C - C$ and $C - B$ bonds \cite{PhysRevB.77.064518}. These effects, associated with structural disorder, have yet to be studied theoretically and applied to BDD or BNCD.

\par{In this study, we address the role of structural and on-site disorder within the narrow boron acceptor level. We consider a triangular lattice with a single boron acceptor band employing the inhomogeneous BdG theory. Disorder is treated in the form of random potential fluctuations at the boron sites as well as structural disorder in the form of a non-uniform hopping integral in the tight-binding term between adjacent sites. The level of structural disorder is correlated to the atomic boron concentration as this disorder is expected to increase due to local changes in bond length between $C - B$ bonds and surroungding $C-C$ bonds. We include the Coulomb interaction between holes in the acceptor band as it is a narrow band so the effect of the Coulomb interaction needs to be recknoned with \cite{Baskaran}. We study the local pairing amplitude as well as the variation of the mean pairing amplitude, spectral gap and superfluid density as the disorder level increases. Through interpretation of the density of states, mean pairing amplitude, spectral gap, local superfluid density and phase stiffness we isolate the differences between structural and on-site disorder. We illustrate some overlap with experimental work, highlighting features of superconducting BDD and BNCD which can be explained through the inclusion of structural disorder.} 

\section{Model}

We solve a tight-binding model considering a triangular lattice, incorporating the Coulomb interaction between acceptor states within the attractive Hubbard model. We employ the mean field approximation and generate the Hamiltonian using the BdG theory. The Hamiltonian is given by

\begin{eqnarray}
{\cal H}=&& \sum_{  i  ,\sigma} (\epsilon_{i} - \mu_{i}) c_{i \sigma}^{\dagger} c_{i \sigma} +  \sum_{i} (\Delta (r_{i}) c_{ i\uparrow}^{\dagger} c_{i \downarrow}^{\dagger} + \Delta^{*} (r_{i}) c_{ i\downarrow} c_{i \uparrow}) \nonumber\\
&&-  \sum_{\langle i,j \rangle , \sigma} t_{ ij } (c_{i \sigma}^{\dagger} c_{j \sigma} + h.c.) + \sum_{\langle i,j \rangle} W_{ij} (n_{i}n_{j}),
\end{eqnarray}

where $\epsilon_{i}$ is the potential at each site, $\mu_{i}$ is the chemical potential at each site with $c_{i \sigma}^{\dagger}$ ($c_{i \sigma}$) being the fermion operators of creation (annihilation) at each site. The hopping integral between adjacent sites is $t_{ij}$ while the strength of the Coulomb interaction between adjacent sites is $W_{ij}$. $ \Delta(r_{i}) $ is the local order parameter with co-ordinate $ r_{i} $ at site $ i $ and is given by $\Delta (r_i) = -U \langle c_{i \uparrow} c_{i \downarrow} \rangle$ where $U$ is the interaction strength. 

This Hamiltonian is rendered diagonalizable through the Bogoliubov-Valatin transformations for the fermion operators

\begin{equation}
c_{i \sigma}(r) = \sum_{i, \sigma} u_{n \sigma \sigma'} (r) \gamma_{n \sigma'} + v_{n \sigma \sigma'}^{*}( r) \gamma_{n \sigma'}^{\dagger} 
\end{equation}

The local order parameter can then be expressed as 

\begin{eqnarray}
\Delta(r_{i})=&& -U \sum_{n} u_{n \uparrow} (r_{i}) v_{n \downarrow}^{*}(r_{i}) [ 1 - f(E_{n}) ] \nonumber\\
&&+ u_{n \downarrow}(r_{i}) v_{n \uparrow}^{*}(r_{i})f(E_{n}),
\end{eqnarray}

where $ f(E_{n}) $ is the Fermi-Dirac distribution function. The local occupation number is given by

\begin{equation}
n(r_{i}) = \sum_{n} \vert u_{n \uparrow} (r_{i}) \vert^{2} f(E_{n}) + \vert v_{n \uparrow} (r_{i}) \vert^{2} (1 - f(E_{n}))
\end{equation}

The resulting Hamiltonian is of the form

\begin{equation}
\sum_{j} \left(
	\begin{matrix}
	{\cal H}_{i,j} & \Delta_{i,j} \\
	\Delta_{i,j}^{*} & -{\cal H}_{i,j}^{*}
	\end{matrix} \right) 
	\left(
	\begin{matrix}
	u_{n}(r_{j}) \\
	v_{n}(r_{j})
	\end{matrix}
	\right)
	=
	E_{n} 	\left(
	\begin{matrix}
	u_{n}(r_{j}) \\
	v_{n}(r_{j})
	\end{matrix}
	\right)
\end{equation}

\par{which is diagonalized self-consistently until sufficient convergence of the order parameter. We work with a triangular lattice of  $ 25 \times 25 $ sites, having found this to be sufficiently large to overcome finite size effects. All energy scales were normalized to the hopping integral. The chemical potential of the boron sites (relative to the carbon sites) was determined by calculating self-consistently the chemical potential which would give the correct occupation. }

The local density of states is given by 

\begin{eqnarray}
\rho(E) = \frac{1}{N} \sum_{n,r_{i},\sigma}  \vert u_{n \sigma} (r_{i}) \vert^{2} \delta (E - E_{n})
\nonumber\\
+ \vert v_{n \sigma } (r_{i}) \vert ^{2} \delta (E + E_{n}) 
 \label{doseqn}
\end{eqnarray}

where $ N $ is the total number of lattice sites.

\par{The superfluid density is calculated to study the rigidity of the superconducting phase \cite{Das,Huang,BC}. This is achieved through the linear response of the system to an external time-dependent vector potential applied in the $ x $ direction, $ A_{x} (r,t) = A(q, \omega) e^{i \mathbf{q} \cdot \mathbf{r}_{i} - i \omega t}$. The hopping term is then modified by the Peierls phase factor, $ c_{i + x , \sigma}^{\dagger} c_{i \sigma} \rightarrow c_{i + x , \sigma}^{\dagger} c_{i \sigma} e^{ieA_{ij}} $ where $A_{ij} = \int^{\mathbf{r}_{i}}_{\mathbf{r}_{j}} \mathbf{A}(\mathbf{r},t) \cdot d \mathbf{r} $ (with units $c = 1 = \hbar$) and $i + x$ represents the next site in the $x$ direction and $e$ is the electron charge . These terms are expanded to second order in $A$ yielding

\begin{equation}
H'(t) = - \sum_{i} \left( e j_{x}^{p}(\mathbf{r}_{i}) A_{x}(\mathbf{r}_{i},t) + \frac{e^{2} k_{x}(\mathbf{r}_{i})}{2} A^{2}_{x} (\mathbf{r}_{i},t) \right)
\label{hprime}
\end{equation}

with a  kinetic energy density (diamagnetic response)

\begin{equation}
k_{x}(\mathbf{r}_{i}) = - \sum_{\sigma} ( t_{i,i+x} c^{\dagger}_{i \sigma} c_{i + x, \sigma} + H.c. )
\label{k}
\end{equation}

and paramagnetic current density
 
 \begin{equation}
j^{p}_{x} ( \mathbf{r}_{i} ) = -i \sum_{\sigma} (t_{i,i+x} c^{\dagger}_{i \sigma} c_{i+x, \sigma} + H.c.) \,.
\label{sfdexp}
\end{equation}

The charge current density operator is given by the functional derivative 

\begin{equation}
J_{x}^{Q}(\mathbf{r}_{i}) = - \frac{\delta H'(t)}{\delta A_{x}(\mathbf{r}_{i},t)} \,.
\label{ccd}
\end{equation}

The superfluid weight $\rho_{s}$, defined as the superfluid density divided by mass, is then given by

\begin{equation}
\frac{\rho_{s}}{m^{*}} = - \frac{\langle J_{x}^{Q}(r_{i}) \rangle }{e^{2} A_{x}(r_{i})}	 \,.
\label{sfdhuang}
\end{equation}

The average over the paramagnetic current density can be expressed, in the interaction representation, as 

\begin{equation}
\langle j^{p}_{x} (r_{i}) \rangle = -i \int_{- \infty}^{t} \langle [j^{p}_{x} (r_{i},t), H'(t')] \rangle dt' \,.
\label{pmcdavg}
\end{equation}
 
The long-wavelength, zero frequency limit yields the superfluid weight in the direct current regime which is studied here. Using the relation $ H' = - \int d^{3}r j^{p}_{x}(r_{i}) A(\mathbf{r},t) $ the thermodynamic average over the paramagnetic current density can be reduced to 

\begin{equation}
\langle j^{p}_{x}(r_{i}) \rangle = - \frac{e A_{x}(r,t)}{N} \Pi_{xx}(\mathbf{q}, \omega) \,.
\label{jpavg}
\end{equation}

This is evaluated in the Matsubara formalism as a current-current correlation function  

\begin{equation}
\Pi_{xx}(\mathbf{q},iw_{n}) = - \int_{0}^{1/k_{B}T} d \tau e^{iw_{n} \tau} \langle T_{\tau} j^{p}_{x}(\mathbf{q} , \tau) j^{p}_{x}(-\mathbf{q} , 0) \rangle
\label{cccor}
\end{equation}

Finally, the superfluid weight can be expressed as

\begin{equation}
\frac{\rho_{s}}{m^{*}} = \langle -k_{x} \rangle - \Pi_{xx}(q_{x} = 0, \omega = 0) \,.
\label{sfweight}
\end{equation}

\subsection{The incorporation of disorder}

\par{We study two broad classes of disorder. Initially, we consider disorder in the form of random deviations in the local on-site potential energy at the boron sites only. We assume a normal distribution about some mean value which corresponds to the correct chemical potential to set the correct occupation. The full width at half maximum of the normal distribution is referred to as the disorder parameter, $ \sigma $. The disorder parameter is quoted in terms of the tight binding hopping parameter, $t$, as all energies are normalized to this scale. We also study the case of random on-site disorder at each lattice site in order to directly compare the characteristics of on-site disorder to those of structural disorder.}

\par{We then study the influence of structural disorder. We assume that bond length disorder results in changes in the tight-binding hopping integral through a deformation potential \cite{mkrmsb} (which relates the change in distance between atomic sites to a change in the hopping integral between the sites). Structural disorder is also incorporated through random nearest neighbour hopping parameters ($ t_{i,j} $) following a normal distribution about some mean value with full width at half maximum of $ \sigma $. We assume structural disorder throughout all lattice sites. In the case of BNCD grains, structural disorder will likely play a large role due to the inherent structural inhomogeneity.}

\par{We also study a form of correlated structural disorder where the structural disorder parameter changes as the boron atomic concentration changes. Experimental studies have shown that boron incorporation in diamond and nanodiamond is accompanied by microstructural changes in the films due to slight modification of the local bonding invironment in the region of the boron impurities \cite{Kawano2010}. At this stage, we assume three different forms of correlated disorder i.e. disorder parameter increasing linearly with the atomic boron concentration, increasing exponentially and inverse exponential increase.}


\section{Results}

\subsection{On-site potential disorder}

\begin{figure}[!ht]
\centering
\epsfig{file=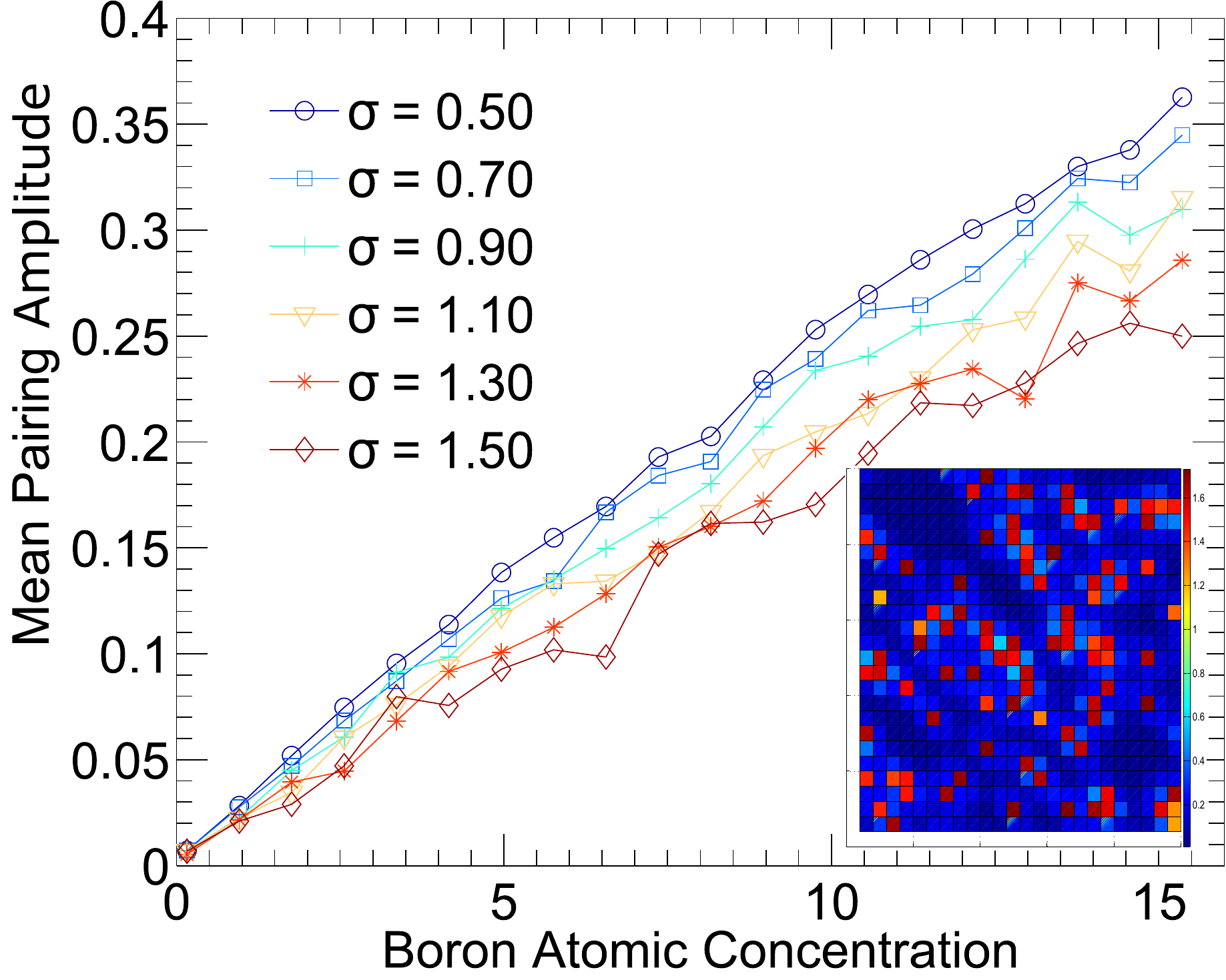,width=85mm}
\caption{Mean pairing amplitude as a function of the boron concentration considering only on-site disorder where the disorder parameter is varied from 0.5 to 1.5. Inset is the local pairing amplitude considering a small on-site disorder parameter $\sigma = 0.05$}	
\label{mpg_vs_b_highdisorder}
\end{figure}

\par{Although the upper limit of the boron doping concentration in diamond is relatively low (up to around 5 \%), we focus on doping concentrations from 0 \% up to 15 \% to highlight the effects of disorder. The inset in Fig. \ref{mpg_vs_b_highdisorder} shows the local pairing amplitude at each lattice site where there is only on-site disorder considering a small disorder parameter (0.05). The local pairing amplitude is finite only at or close to the boron sites and correlates well with the local occupation in the low disorder regime (not shown here).}


\par{Figure \ref{mpg_vs_b_highdisorder} shows the mean pairing amplitude as a function of the atomic boron concentration for various on-site disorder parameters from 0.05 to 1.5 considering on-site disorder only at the boron sites. The mean pairing amplitude varies non-smoothly with increasing disorder parameter. Figure \ref{mpg_vs_b_highdisorder} shows that while on-site disorder can suppress the mean pairing amplitude to some extent, this is only possible if the potential fluctuations are very large, on the order of the hopping parameter. This indicates that disorder in the form of on-site energy of the boron impurities alone is not sufficient to reproduce the experimentally reported suppression (at relatively low doping concentrations) of the pairing amplitude with increasing boron concentration. Comparison of the local on-site energy with the local pairing amplitude (not shown here) indicates that where the on-site energy is reduced, the pairing amplitude increases.}

\begin{figure}[!ht]
\centering
\subfloat{\epsfig{file=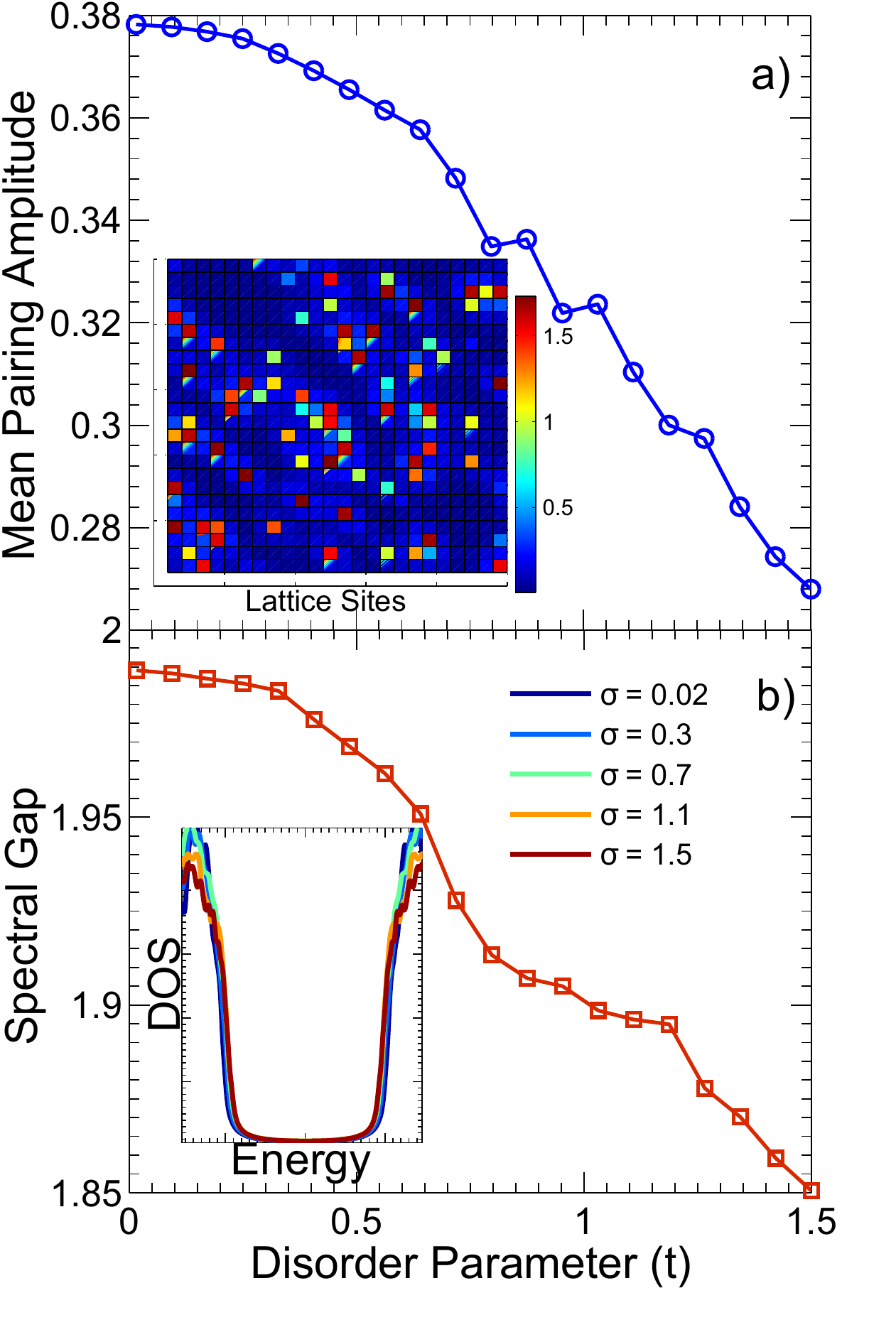,width=70mm}}\\
\subfloat{\label{dossublab}\epsfig{file=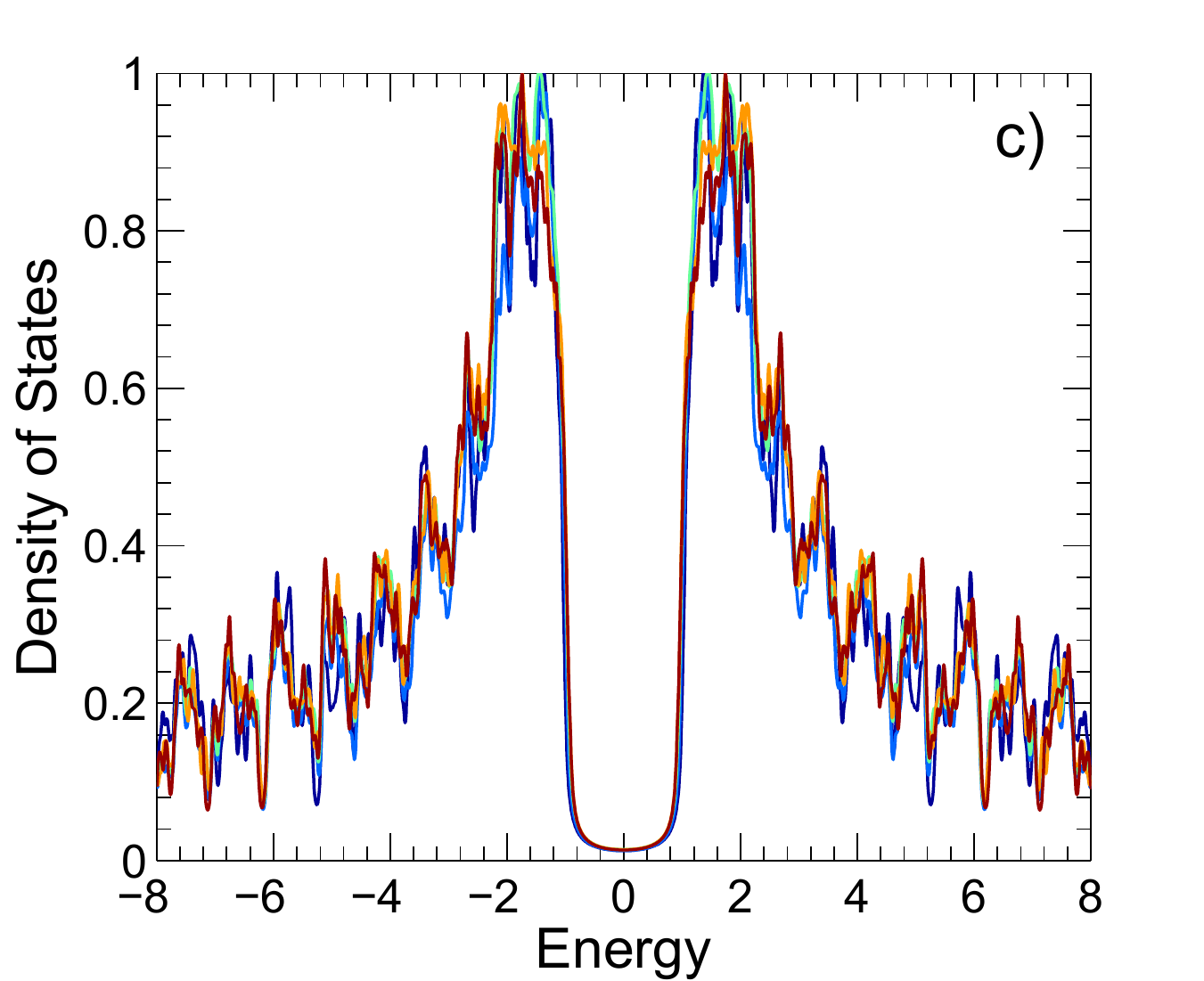,width=70mm}}
\caption{a) Mean pairing amplitude as a fucntion of the on-site disorder paramater (considering only on-site potential disorder at the boron sites only) where the atomic boron concentration is held fixed at 15 \%. Inset is the local pairing amplitude at each lattice site with a relatively large on-site disorder parameter of $\sigma = 1.5$.  b) The corresponding spectral gap as a function of the on-site disorder paramater. Inset is the density of states around the gap. Inset is the density of the states in the vicinity of the gap. c) Density of states where the boron concentration is held fixed and the structural disorder parameter is varied.}	
\label{epsdis_comb}
\end{figure}


\par{To determine the effects of on-site disorder we study the mean pairing amplitude, spectral gap, density of states and the distribution of pairing amplitudes for different disorder parameters while the boron concentration is held fixed, initially assuming on-site potential fluctuations at the boron sites only. The on-site disorder parameter ($ \sigma $) is varied from 0.5 to 1.5. Figure \ref{epsdis_comb}(a) shows the variation of the mean pairing amplitude as the on-site disorder parameter is varied. The mean pairing amplitude decreases with increasing disorder. In the high disorder regime there is a wider distribution of local pairing amplitudes (inset of Fig. \ref{epsdis_comb}(a)).}

\par{The spectral gap decreases with increasing disorder however the overall change is not very large, as illustrated in Fig. \ref{epsdis_comb}(b). The spectral gap follows a similar trend to the mean pairing amplitude. The variation is inherently noisy as we have assumed a very small smearing factor (to approximate the delta function, see Eq. \ref{doseqn}) when calculating the DOS so calculations of the spectral gap are very sensitive to the very small changes in the disorder parameter which arise due to on-site disorder).}

\par{The density of states (Fig. \ref{epsdis_comb}(c)) shows very little change, despite the on-site disorder parameter varying largely. The spectral gap remains intact and does not transition to a pseudo-gap however there is a small decrease in the density of states in the region close to the gap (corresponding to a decrease in the sharpness of the coherence peaks). The noise in the DOS is due to the small smearing factor we have used in approximating the delta function in Eq. \ref{doseqn} which is necessary to reveal the small changes which take place in the vicinity of the gap.}

\par{Figure \ref{histsublab} shows the distribution of the local pairing amplitudes considering only on-site disorder at the boron sites. When the disorder parameter is low the number of lattice sites with an almost uniform pairing amplitude of  around 1.6$t$ is relatively large while most lattice sites have zero pairing amplitude. With increasing on-site disorder, the number of lattice sites with pairing amplitude around 1.6$t$ decreases while the number of lattice sites with pairing amplitude around zero increases. Except for relatively high disorder (disorder parameter 1.5) there are few lattice sites with pairing amplitude between 0.6$t$ and 1.2$t$. We study the correlation functions of the local pairing amplitude, $ D(|r_{i} - r_{j}|) = \langle \Delta(r_{i}) \Delta(r_{j}) \rangle $ to glean insight into the range of the order parameter. The inset in Fig. \ref{histsublab} shows the pairing amplitude correlation function considering only on-site disorder with a small disorder parameter ($0.05 t$, bule curve) while the red curve shows the corresponding correlation function where the structural disorder parameter is larger ($1.5 t$). The change in the correlation functions is not large however there is a crossover in the correlation functions as the distance increases.}

\begin{figure}[!ht]
\centering
\epsfig{file=
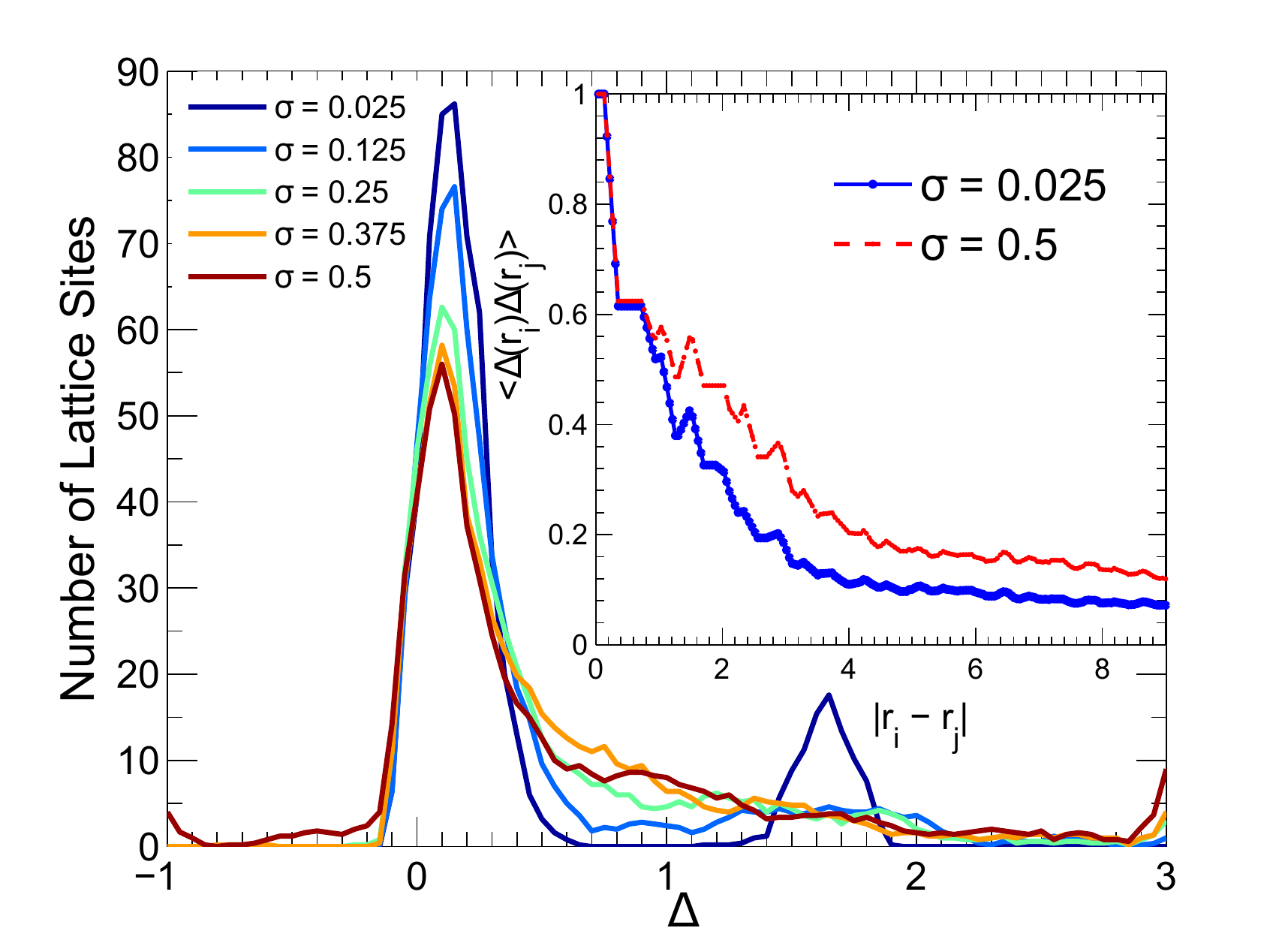,width=70mm}
\caption{Distribution of the local pairing amplitude for different realizations of  the on-site disorder parameter considering only on-site disorder at the boron sites. Inset is the pairing amplitude correlation function considering only on-site disorder with a disorder parameter of $\sigma = 0.02$ (blue) and $\sigma = 1.5$ (red).}	
\label{histsublab}
\end{figure}



\begin{figure}[!ht]
\centering
\epsfig{file=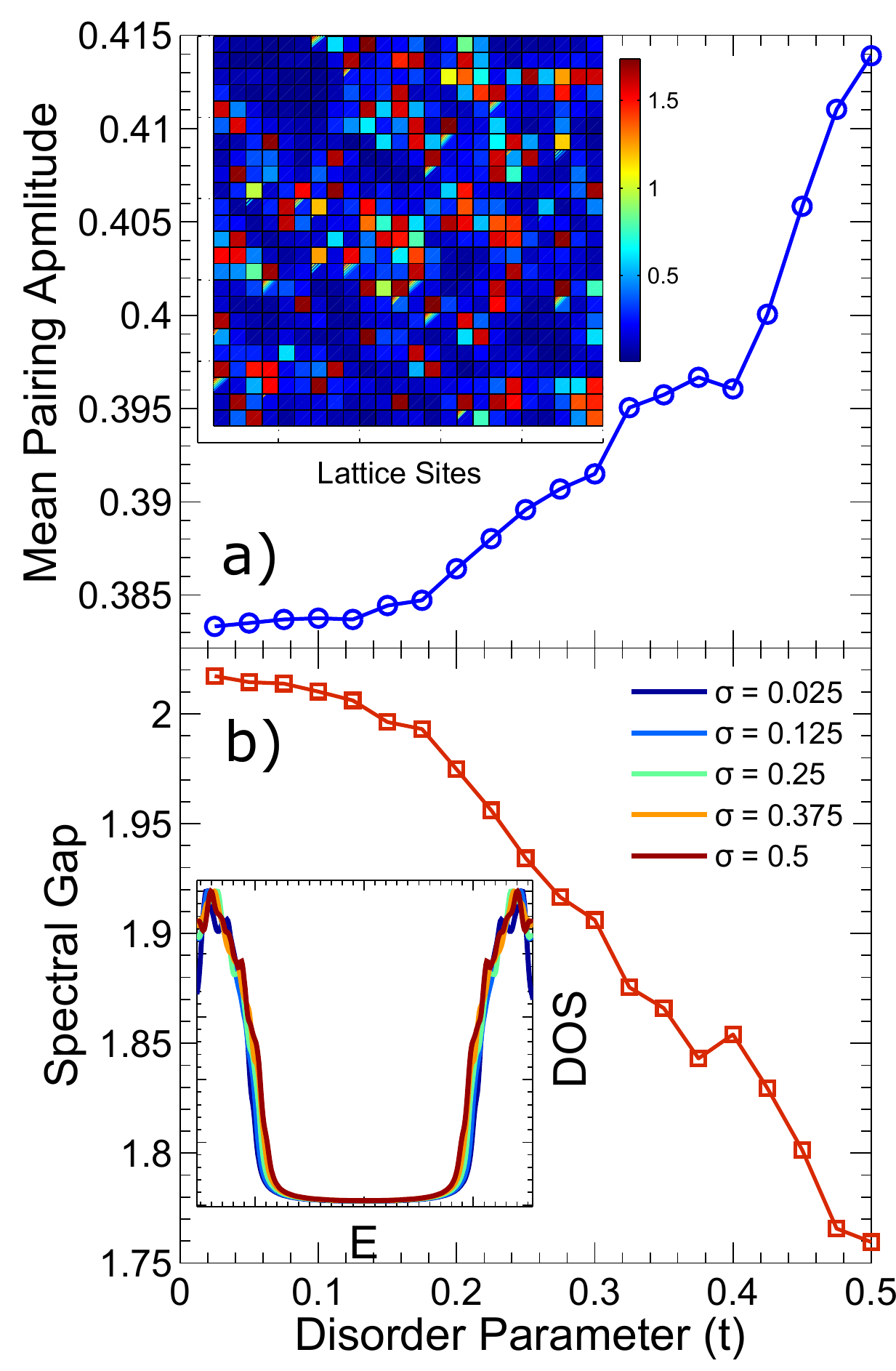,width=70mm}
\caption{a) Mean pairing amplitude as a function of the on-site disorder parameter considering on-site disorder at all sites in the lattice at a constant atomic boron concentration of 15 \%. Inset is the local pairing amplitude considering an on-site disorder parameter of $\sigma = 1.5$.	b) The corresponding spectral gap. Inset is the density of states around the gap region.}	
\label{epsdisALL_comb}
\end{figure}


\begin{figure}[!ht]
\centering
\epsfig{file=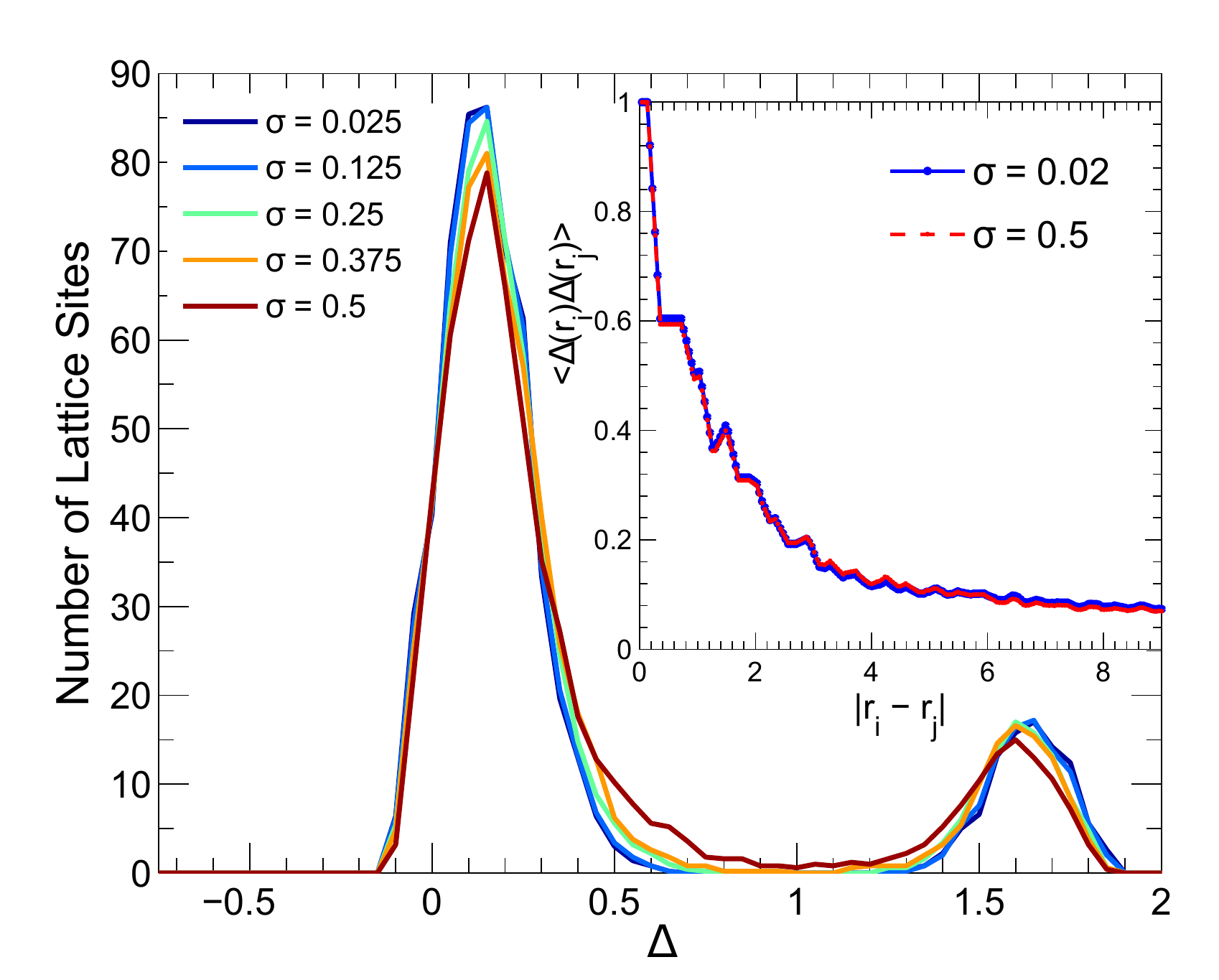,width=70mm}
\caption{Distribution of the local pairing amplitude as a function of the boron concentration for different realizations of the on-site disorder parameter considering on-site disorder at all sites. Inset is the pairing amplitude correlation function considering a small on-site disorder parameter (blue dots) and a larger diosrder parameter (red dashed line).}	
\label{epsALLspec_gap}
\end{figure}


\par{We then assume that there are on-site potential fluctuations at all sites in the lattice, as opposed to only where boron is incorporated. Figure \ref{epsdisALL_comb}(a) shows the variation of the mean pairing amplitude with on-site disorder while the boron concentration held at 15 \%. In this case, the mean pairing amplitude increases slightly as the disorder parameter increases, in contrast to the case where on-site potential fluctuations are assumed only at the boron sites. The spectral gap however decreases slightly as the disorder parameter increases as shown in Fig. \ref{epsdisALL_comb}(b). Figure \ref{epsALLspec_gap} shows the distribution of local pairing amplitudes where on-site disorder is assumed at all lattice sites (to be compared with the case where on-site disorder is assumed only at the boron sites, Fig. \ref{histsublab}). In this case, the number of sites with local pairing amplitude around zero decreases with increasing disorder (in contrast to the case where there is on-site disorder only at the boron sites) hence the mean pairing amplitude increases. The number of lattice sites with local pairing amplitude around 1.6 decreases slightly. The inset in Fig. \ref{epsALLspec_gap} shows the pairing amplitude correlation function for a small disorder parameter (blue dots) and a large disorder parameter (red dashed line). There is hardly any change at all in the correlation functions.}

\subsection{Correlated Structural disorder}

\par{We now study the effects of structural disorder in the form of a random distribution of the tight-binding hopping parameter (non-diagonal disorder) about a constant mean value for all lattice sites in the absence of on-site disorder. We assume that the structural disorder parameter ($ \sigma $) can increase either linearly, exponentially or inverse exponentially as shown in Fig. \ref{correlated_dis}(a) which shows the variation of the disorder parameter as a function of the boron concentration. We assume these three variations as there is as yet insufficient experimental data directly relating the change in the microstructure of BDD or BNCD to the atomic boron concentration. In the case of nitrogen incorporation in nano-diamond and amorphous carbon films, the variation of the structural disorder with nitrogen incorporation is non-linear \cite{Alibart2010,mkrmsb,Shah2010,Bhat2008} showing an initially rapid increase in disorder and tending towards saturation with increasing nitrogen incorporation. To isolate the effects of structural disorder, we first assume that the on-site energy at all sites is uniform.}

\par{Figure \ref{correlated_dis}(b) shows the variation of the mean pairing amplitude as a function of the boron concentration for the three different variations of correlated disorder. It is evident that structural disorder has a much larger effect than disorder in the form of on-site potential fluctuations (Fig. \ref{mpg_vs_b_highdisorder}). Of particular interest is the case of the disorder parameter increasing inverse exponentially with the boron concentration. The mean pairing amplitude shows significant saturation with the boron concentration, especially in the region of 5 - 6 \% atomic boron concentration (this is more significant than in the case of on-site disorder only) and is similar to experimental results \cite{Kawano2010}.}


\begin{figure}[!ht]
\centering
\epsfig{file=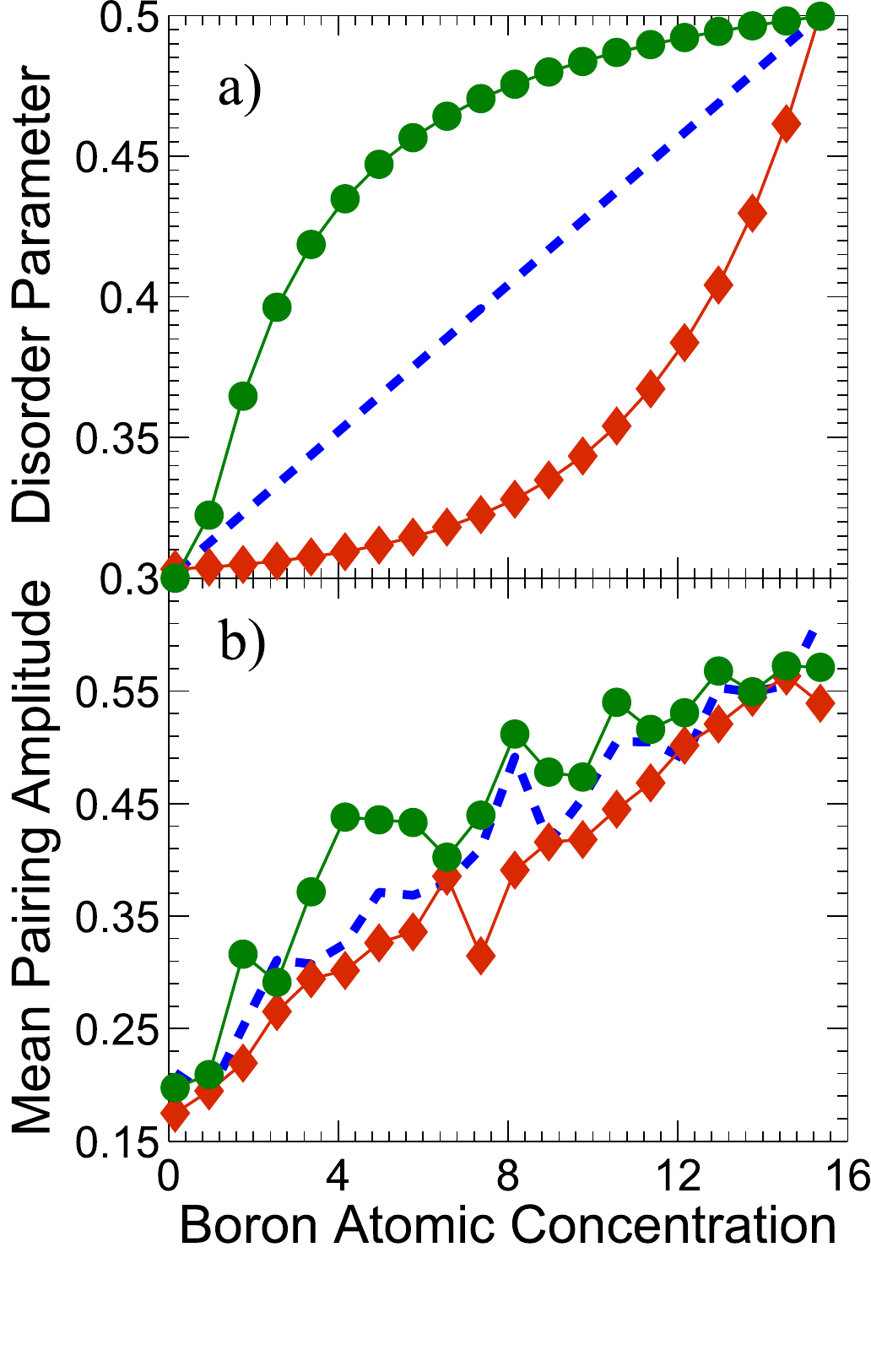,width=70mm}
\caption{a) Variation of the structural disorder parameter as a function of the atomic boron concentration for three different cases. b) The mean pairing amplitude as a function of the atomic boron concentration for three different cases. The case of linear correlated disorder corresponds to the blue dashed line while the expoential variation of disorder corresponds to the red diamonds and inverse exponential variation corresponds to the green dots.}	
\label{correlated_dis}
\end{figure}


\par{To illuminate the influence of structural disorder, we study the mean pairing amplitude, the distribution of pairing amplitudes, the spectral gap and the density of states while the boron concentration is fixed at 15 \% and the structural disorder parameter is varied (assuming no on-site disorder).  Figure \ref{strucdiscom}(a) shows the variation of the mean pairing amplitude as a function of the structural disorder parameter. The mean pairing amplitude shows an unexpected significant increase with increasing structural disorder. This is in contrast to the case of on-site disorder with random fluctuations at the boron sites only (refer to Fig. \ref{epsdis_comb}(a)) where the mean pairing amplitude decreases with disorder. The spectral gap as a function of the disorder parameter is shown in Fig. \ref{strucdiscom}(b). The spectral gap  decreases with increasing disorder and the decrease is far more significant than in the case of on-site disorder (whether it be on-site disorder at the boron sites only or throughout the lattice). Figure \ref{strucdiscom}(c) shows the density of states for different structural disorder parameters, from 0 (blue) to 0.5 (red). The significant effect of structural disorder on the spectral gap is evident as the gap rapidly reduces however up to a structural disorder parameter of 0.5 it does not transition to  a pseudo-gap. States become available within the former gap region as the gap narrows.}

\begin{figure}[!ht]
\centering
\subfloat{\label{struc_dis}\epsfig{file=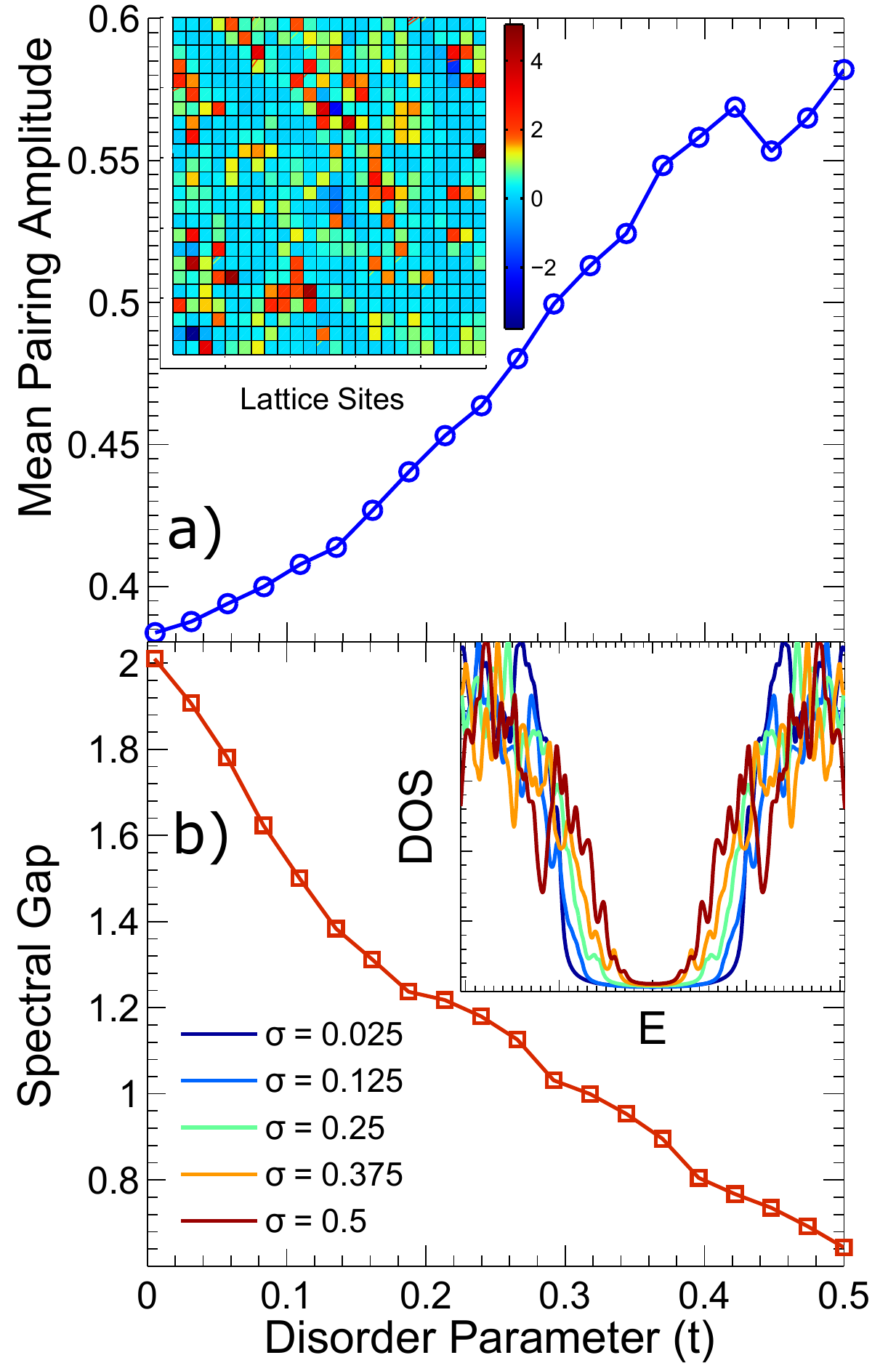,width=70mm}}\\
\subfloat{\label{findos}\epsfig{file=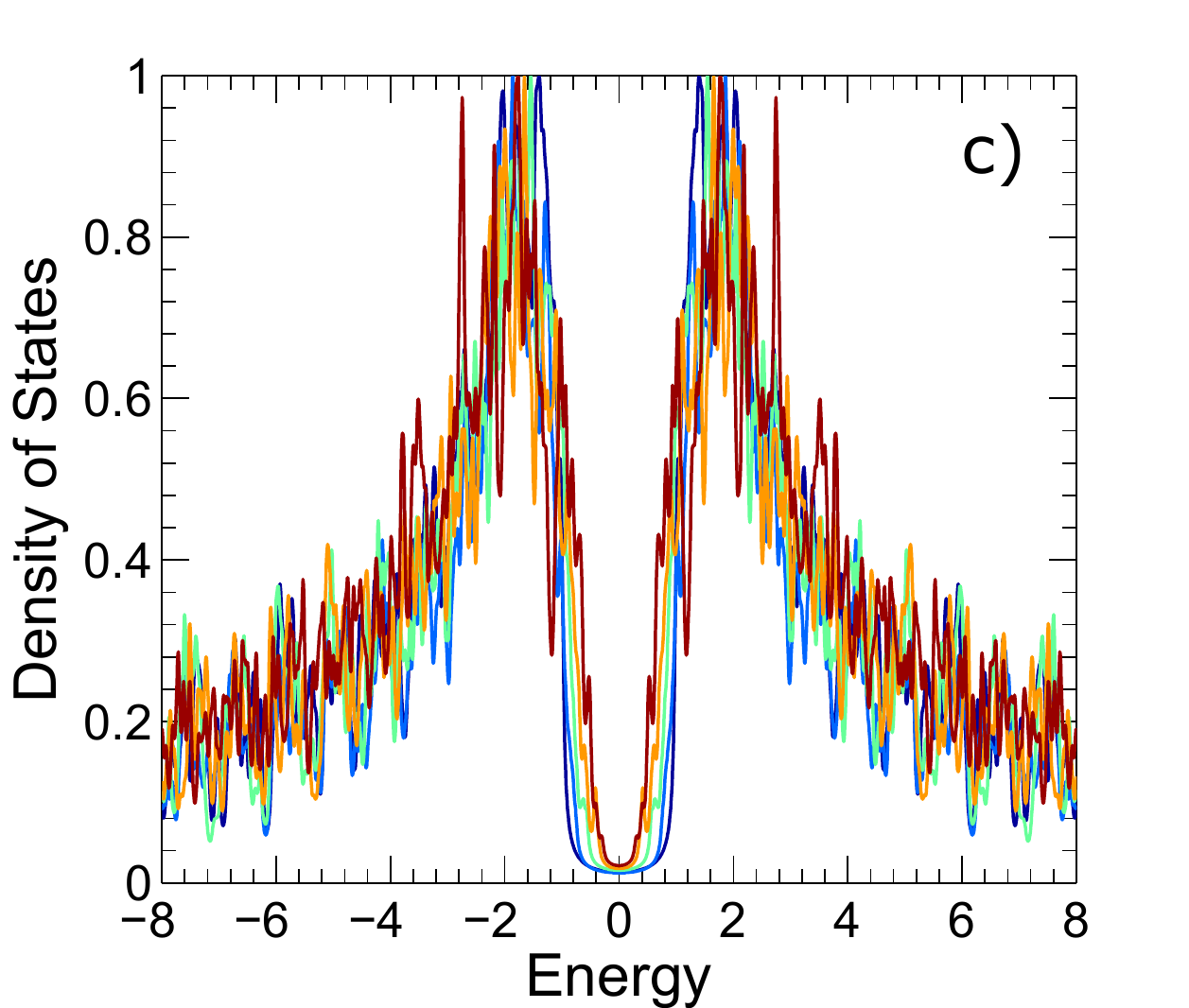,width=70mm}}
\caption{a) Mean pairing amplitude at a constant boron concentration of 15.4 \% while the structural disorder parameter varies from 0 to 0.5 in the absence of on-site disorder. Inset is the local pairing amplitude where the structural disorder patameter is 0.5. b) The corresponding spectral gap as a function of the disorder paramater. Inset is the density of states in the region around the gap. c) The density of states considering different structural disorder parameters.}	
\label{strucdiscom}
\end{figure}


\begin{figure}[!ht]
\centering
\epsfig{file=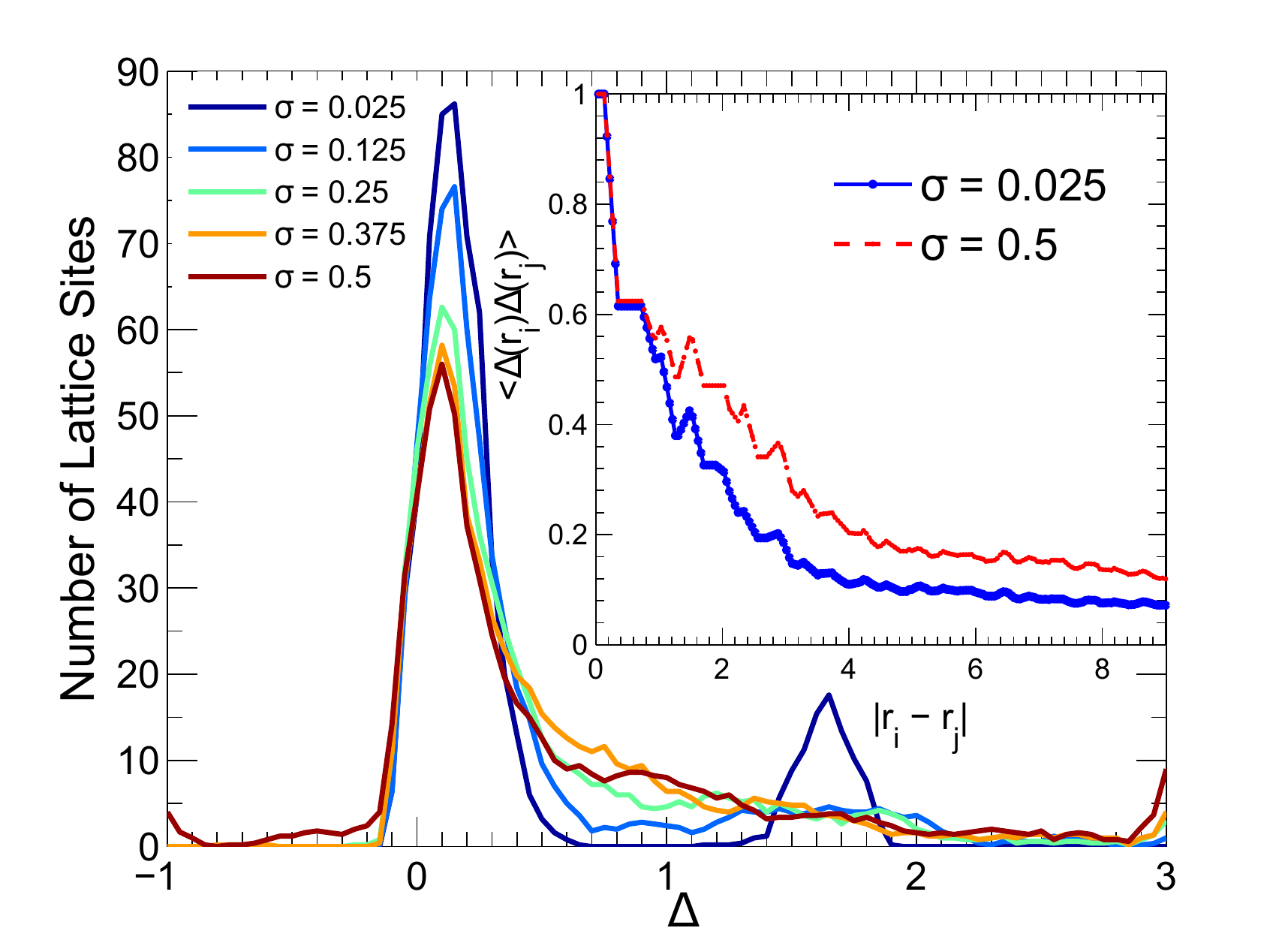,width=70mm}
\caption{Distribution of the local pairing amplitude as a fucntion of the boron concentration for different realizations of structural disorder in the absence of on-site disorder. Inset is the pairing amplitude correlation function for a relatively low disorder parameter (0.025, blue dots) and a higher disorder parameter (0.5, red dashed line).}	
\label{9}
\end{figure}

\par{The distribution of local pairing amplitudes (Fig. \ref{9}) shows that the the number of sites with pairing amplitude in the region of 1.6 diminishes rapidly, in stark contrast to the case of on-site disorder only. On the other hand, the number of sites with zero local pairing amplitude decreases, with the number of sites with non-zero pairing increasing. This shows why the mean pairing amplitude increases overall. In addition, structural disorder promotes connectivity between regions with non-zero pairing. The inset in Fig. \ref{9} shows the pairing amplitude correlation functions for different structural disorder parameters. Correlation persists in the high disorder regime due to the enhanced connectivity of the local pairing amplitude which structural disorder induces.}


\subsection{Combined Structural and On-site Disorder}


\par{In this section structural disorder is combined with on-site disorder while the Coulomb interaction parameter is finite ($W = 3 t$). The variation of the mean pairing amplitude as a function of the boron concentration for different cases of linear, exponential and inverse exponential disorder with boron concentration is shown in Fig. \ref{mpg_comp_all}. The Coulomb interaction parameter decreases the mean pairing amplitude. Small Coulomb interaction parameters do not have a significant influence on the overall mean pairing amplitude . The Coulomb interaction serves to suppress the pairing amplitude as it acts in competition to superconductivity, hence the mean pairing amplitude is reduced. In addition, the overall trend of the mean pairing amplitude as a function of the boron concentration also changes slightly, showing an initial exponential-like increase followed by regions of local maxima.}

\par{Figure \ref{exp_th_comp} shows a comparison of the calculated variation of the mean pairing amplitude along with experimental data from various studies. When the structural disorder parameter varies inverse-exponentially with the boron concentration, the mean pairing amplitude shows clear signs of saturation as reported experimentally \cite{Kawano2010}. Given the energy scales needed to observe saturation of the mean pairing amplitude with boron concentration, this analysis shows that the combination of structural disorder as well as on-site disorder is necessary to reproduce the tendency towards  saturation of the mean pairing amplitude found experimentally. Although the mean pairing amplitude cannot be directly compared to the transition temperature, the variation of the mean pairing amplitude is indicative of the variation of the superconducting transition temperature. In this study we find reasonable qualitative agreement with experiment, finding an initially rapid increase in the mean pairing amplitude followed by a more gradual increase. Assuming that the hopping integral varies in an inverse-exponential manner with increasing boron concentration shows the most prominent change in the gradient with the mean pairing amplitude coming close to saturating as the boron concentration increases. As an extreme case, when the structural disorder parameter is large enough, the spectral gap finally disappears (not shown here) through the combination of structural disorder, on-site disorder and the Coulomb interaction.}

\begin{figure}[!ht]
\centering

\epsfig{file=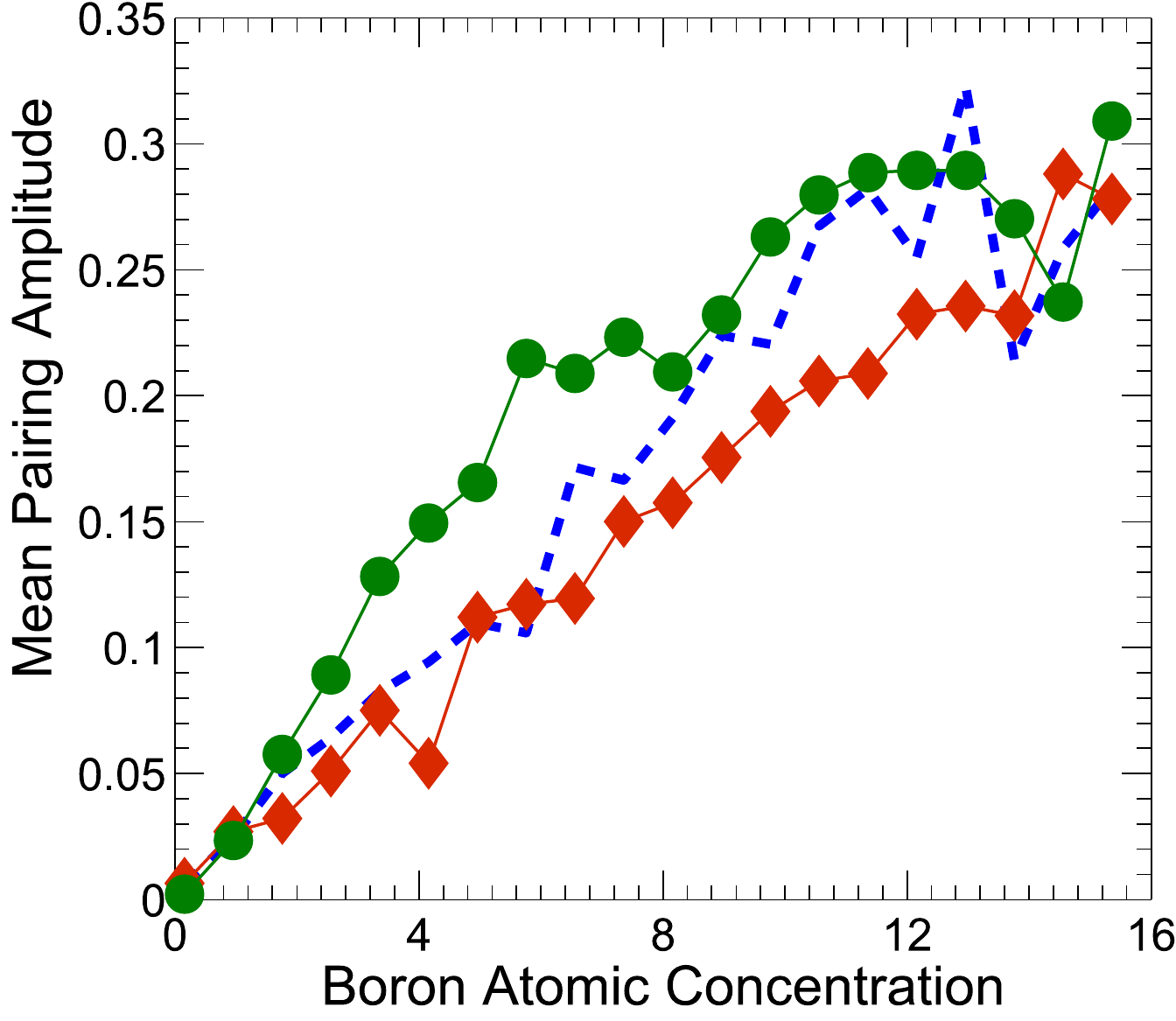,width=70mm}
\caption{Mean pairing amplitude as a function of the atomic boron concentration where the on-site disorder parameter is 0.8, the Coulomb interaction parameter is turned on (W = 3) and the structural disorder parameter varies with the atomic boron concentration. The case of linear correlated disorder corresponds to the dashed blue line while expoentially varying disorder corresponds to the red diamonds and inverse exponentially varying disorder corresponds to the green dots. }	
\label{mpg_comp_all}
\end{figure}

\begin{figure}[!ht]
\centering

\epsfig{file=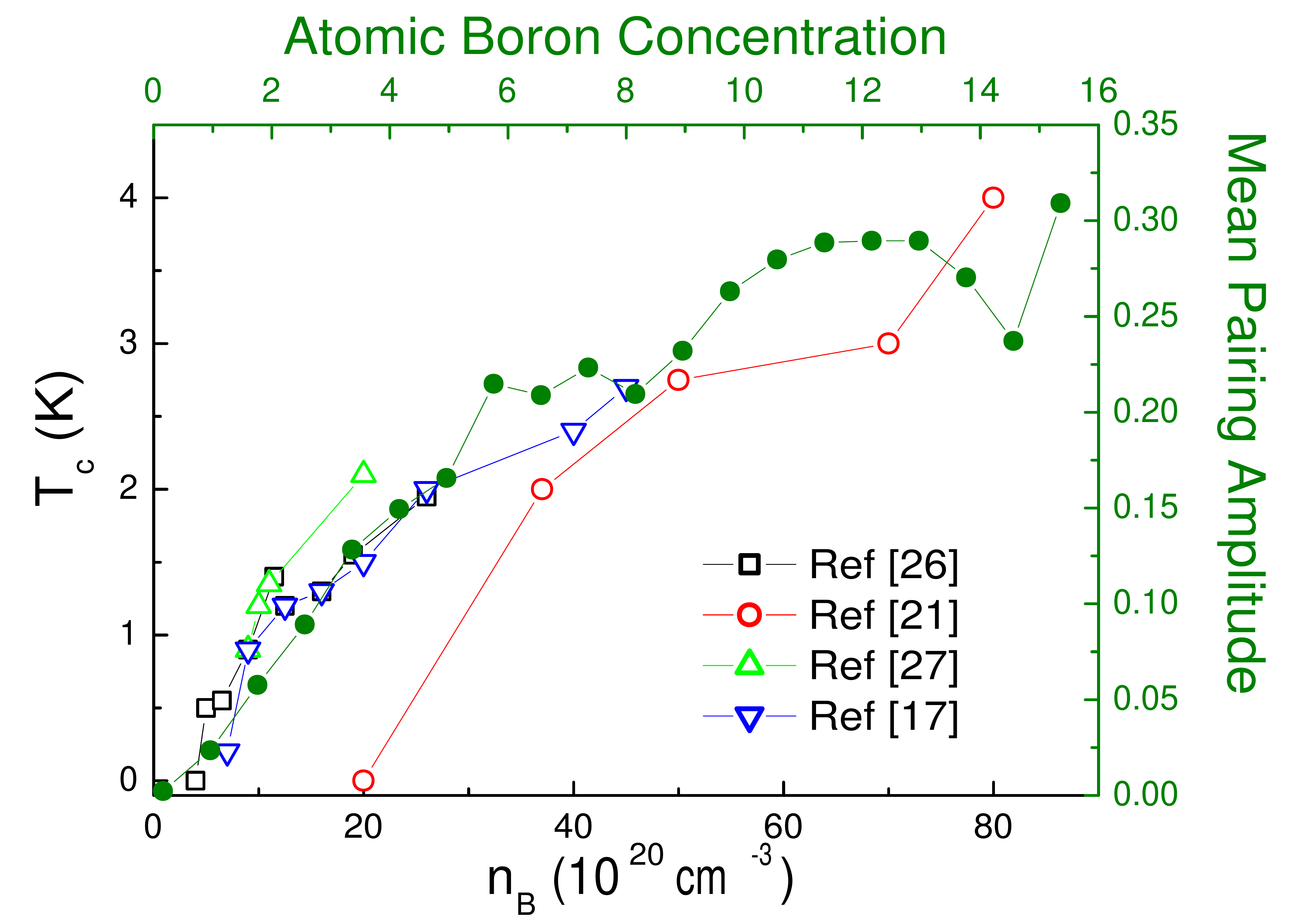,width=90mm}
\caption{Comparison between measured $T_{c}$ as a function of the boron concentration (open symbols) from various references and the calculated variation of the mean pairing amplitude as a function of the atomic boron concentration (filled green circles).}	
	\label{exp_th_comp}
\end{figure}

\begin{figure}[!ht]
\centering

\epsfig{file=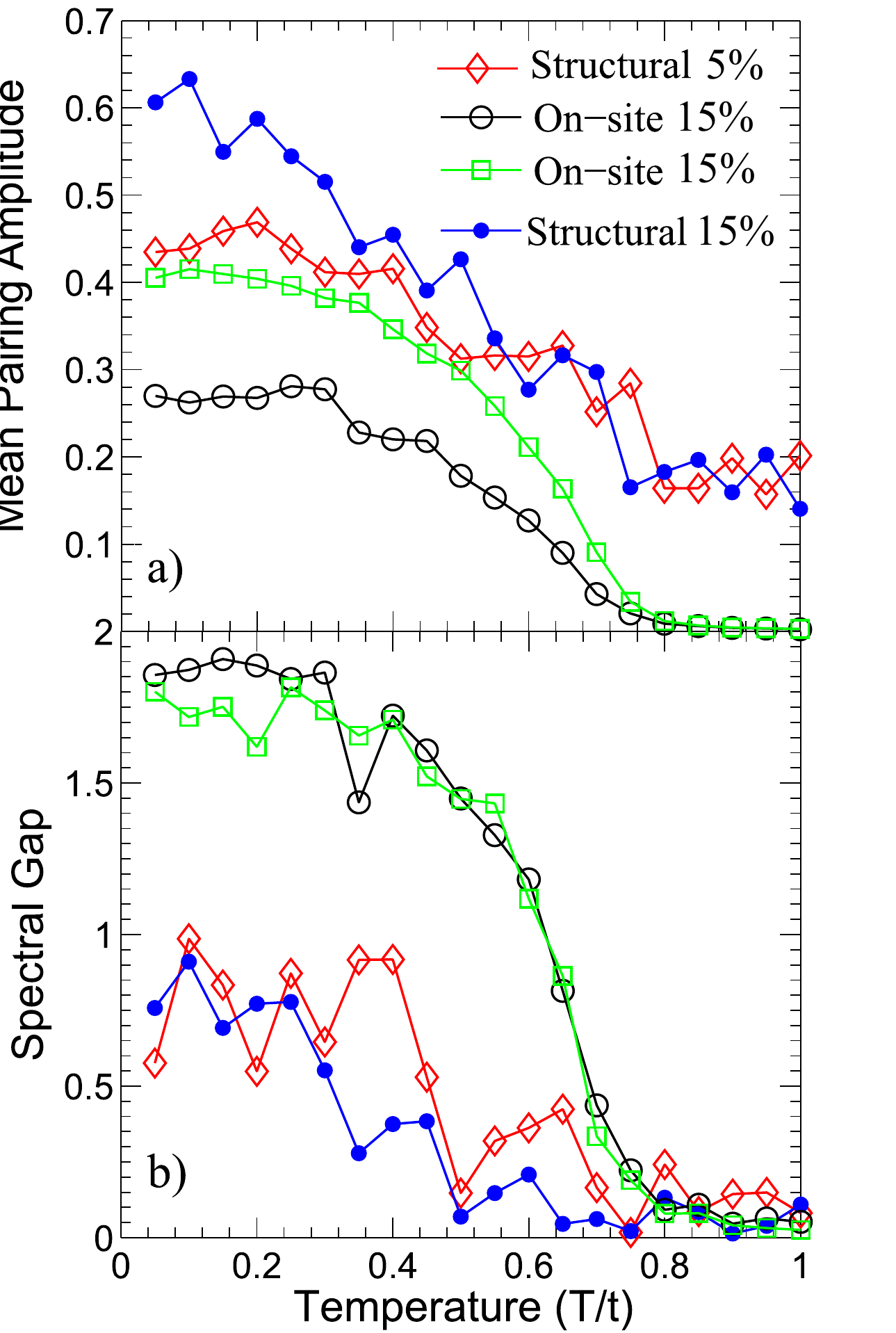,width=70mm}
\caption{The mean pairing amplitude (a) and the spectral gap (b) as a function of temperature for different realizations of disorder. The black open circles represent a high boron atomic concentration (15\%) considering only on-site disorder at the boron sites with a disorder parameter of 1.5 while the green squares represent on-site disorder at all lattice sites with a disorder parameter of 0.5. The red diamonds represent structural disorder with a relatively low boron concentration (5\%) while the blue dots represent structural disorder with a high boron concentration (15\%) while the structural disorder parameter is 0.5.}	
	\label{mpg_TEMP}
\end{figure}

\subsection{Finite Temperature Analysis}

\par{We study some cases of the mean field phase diagram to compare with the experimental phase diagrams. The temperature dependence of the mean pairing amplitude is shown for different cases of both on-site and structural disorder in Fig. \ref{mpg_TEMP}(a). In the case of on-site disorder (whether at the boron sites only (black circles) or at all lattice sites (green squares) with a constant boron concentration of 15 \%) the mean pairing amplitude follows quite closely the behaviour expected for a BCS superconductor even for a relatively large on-site disorder parameter. Considering structural disorder (blue dots represent a high boron concentration, red diamonds a lower boron concentration), the variation of the mean pairing amplitude with temperature is less reminiscent of a standard BCS superconductor. The mean pairing amplitude does not fall to zero within the same temperature range as for on-site disorder (although it does fall to zero at higher temperatures) although the spectral gap (Fig. \ref{mpg_TEMP}(b)) is considerably smaller in the case of structural disorder.}

\begin{figure}[!ht]
\centering

\epsfig{file=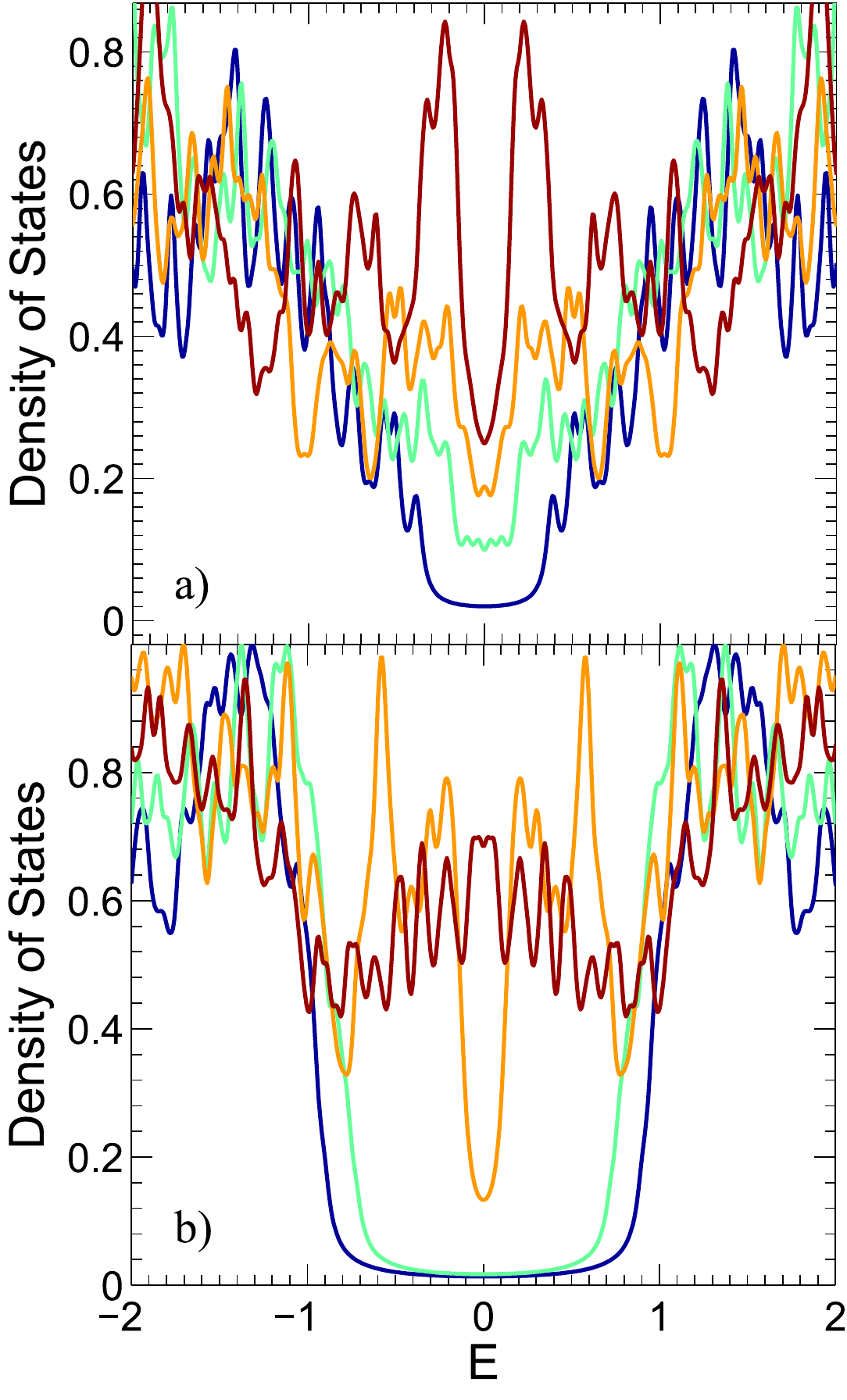,width=70mm}
\caption{Density of states for different temperatures considering only structural disorder (a) and only on-site disorder at all lattice sites (b). The blue curve represents $T = 0$, the green curve $T = 0.5 t$, the orange curve $T = 0.75 t$ and the maroon curve $T = 1.0 t$.}	
	\label{dos_temp_comp}
\end{figure}


\par{To illustrate the fundamental difference between the nature of on-site and structural disorder, we show the DOS at different temperatures from $T = 0$ to $T = t$ considering structural disorder only (Fig. \ref{dos_temp_comp}(a)) and considering only on-site disorder at all sites (Fig. \ref{dos_temp_comp}(b)) for a boron concentration of 15 \% and with the same disorder parameter in both cases. The pseudo-gap which forms with increasing temperature is different for both cases. While the DOS transitions to a deep V pseudo-gap in the case of on-site disorder only, there is an intermediate flat region in the case of structural disorder, which eventually transitions to a deep V with the pseudo-gap persisting to higher temperatures in the case of structural disorder.}

\subsection{Superfluid phase stiffness}

\begin{figure}[!ht]
\centering
\epsfig{file=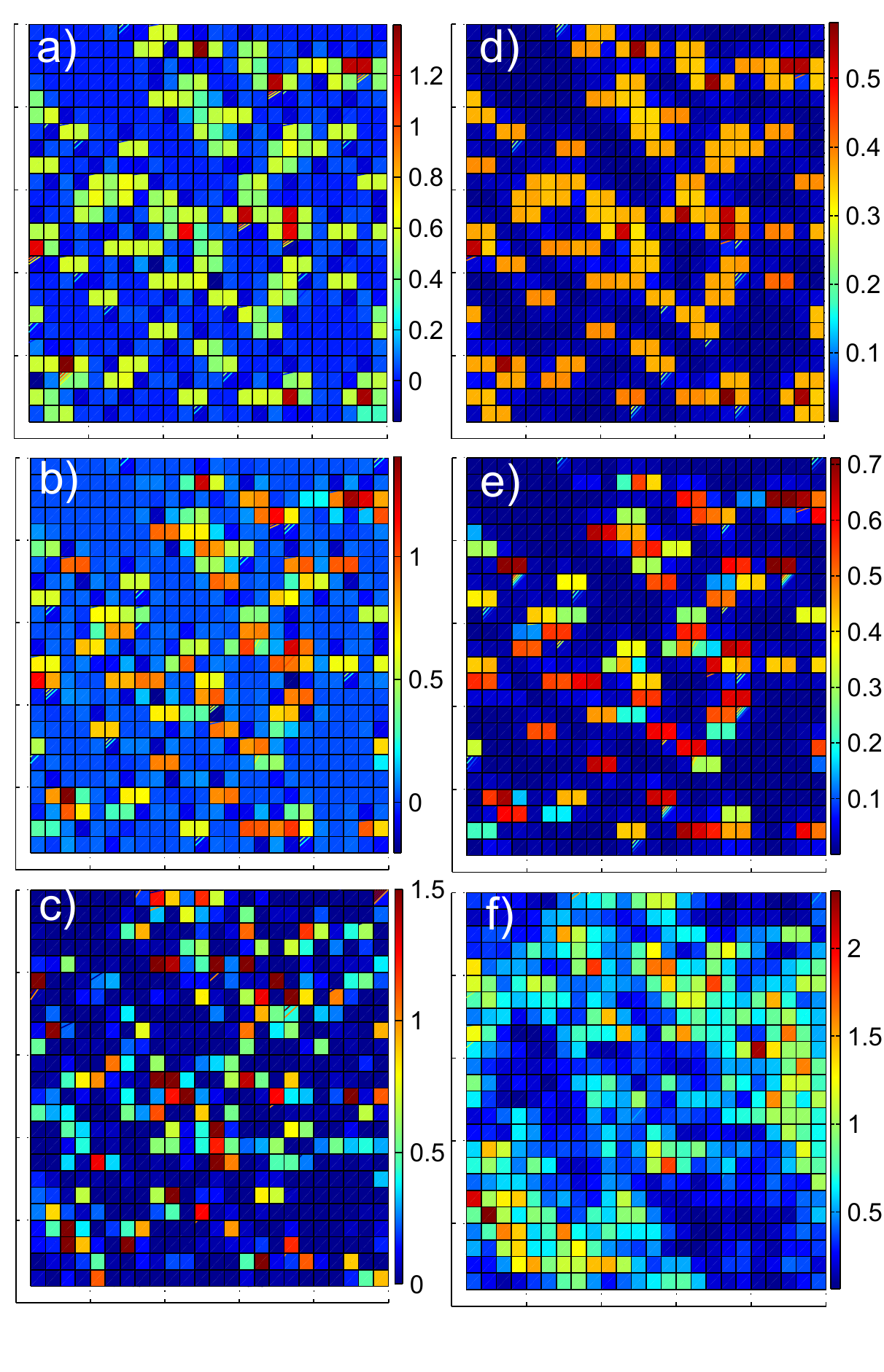,width=85mm}
\caption{a) Local kinetic energy where the on-site disorder parameter is 0.05 considering on-site disorder at the boron sites only. b) Corresponding figure considering on-site disorder at all lattice sites when the disorder parameter is 0.5. c) Local kinetic energy considering only structural disorder where the disorder parameter is 0.5. d) Local paramagnetic response where the on-site disorder parameter is 0.05 considering on-site disorder at the boron sites only. e) Corresponding figure considering on-site disorder at all lattice sites when the disorder parameter is 0.5. f) Local kinetic energy considering only structural disorder where the disorder parameter is 0.5. }
\label{pm_ke_comp}
\end{figure}



\begin{figure}[!ht]
\centering
\subfloat[][]{\epsfig{file=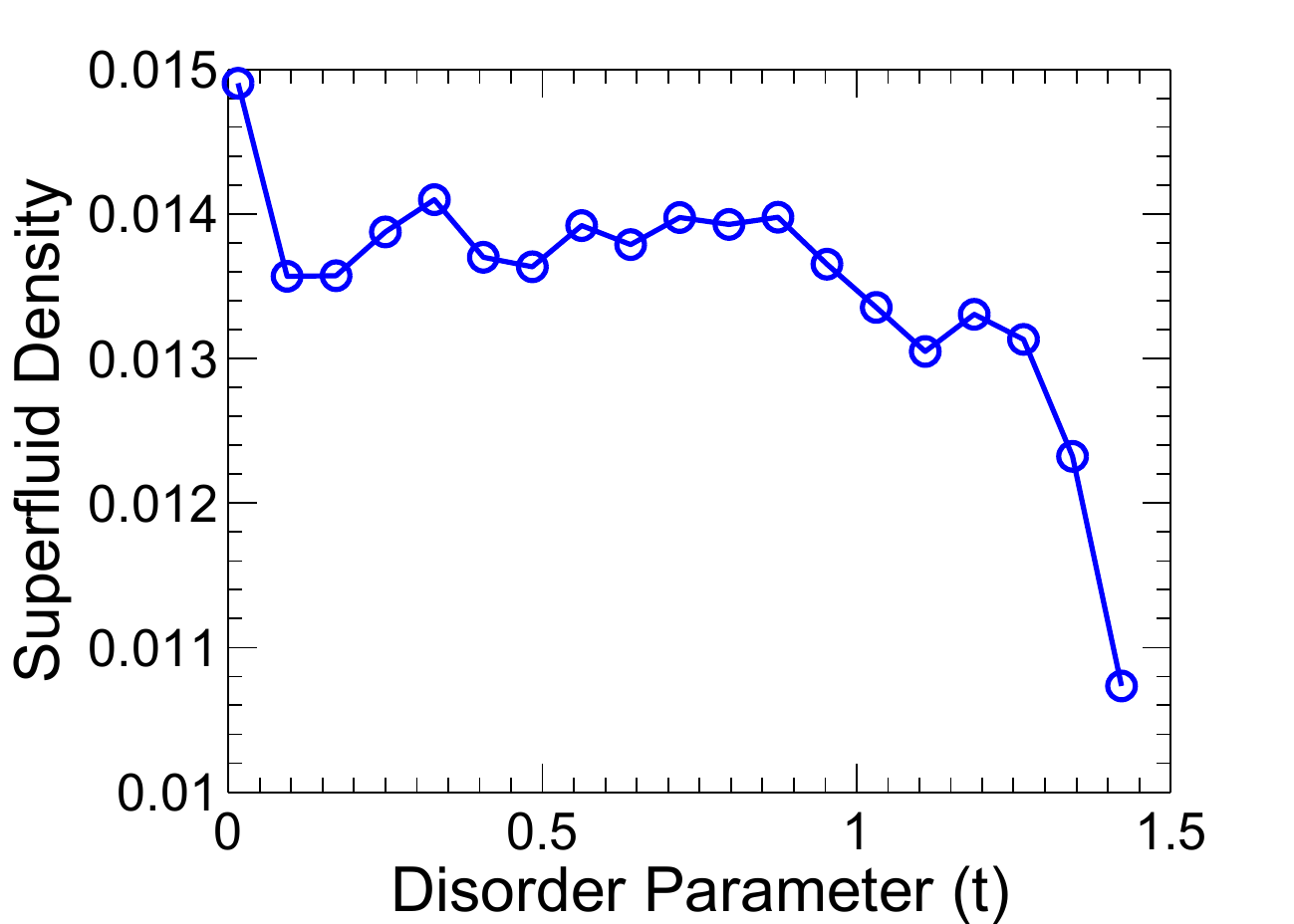,width=60mm}} \\
\subfloat[][]{\epsfig{file=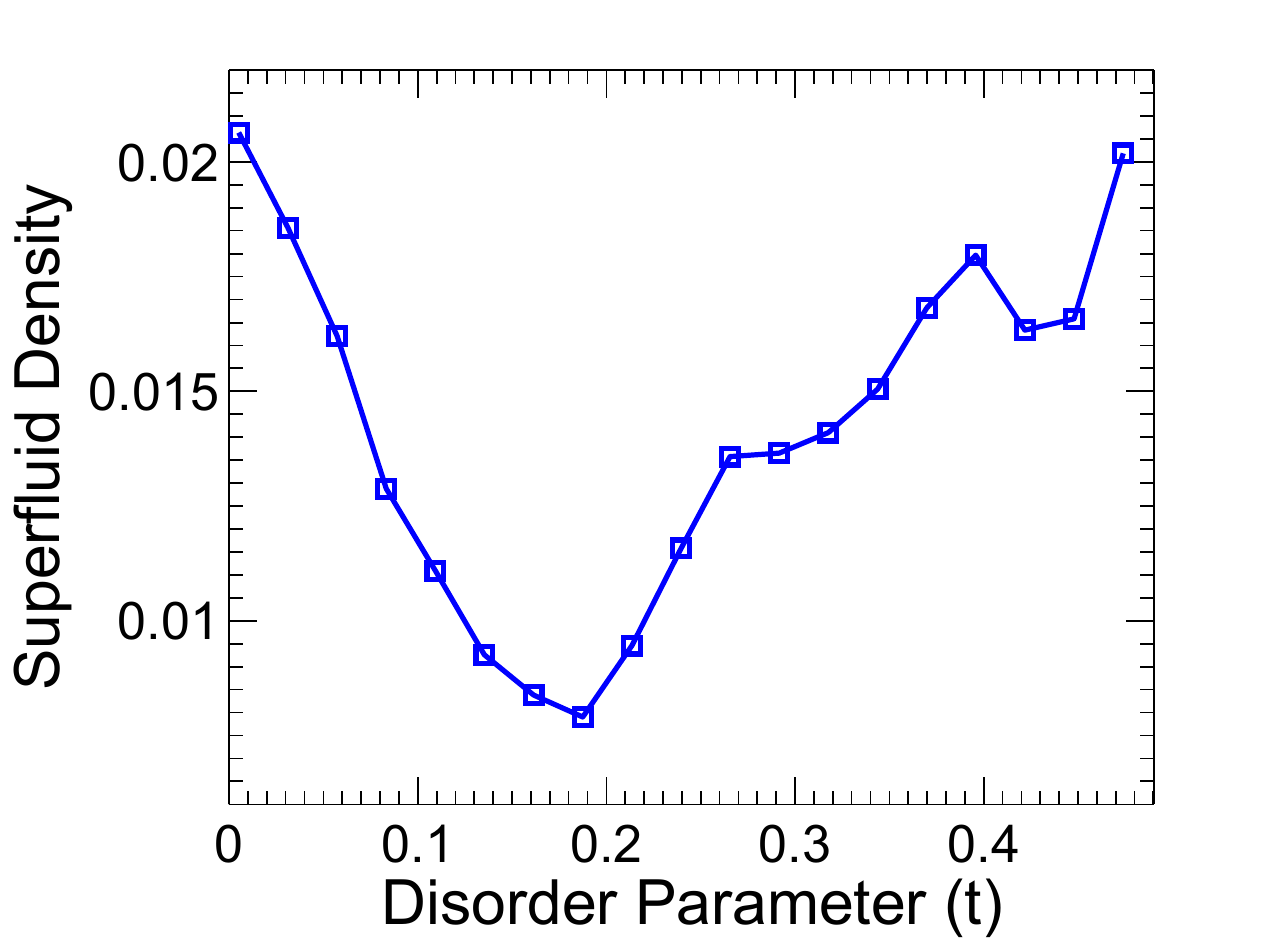,width=60mm}} 
\caption{a) Total superfluid density as a function of the on-site disorder parameter (at a fixed boron concentration). b) Corresponding figure considering structural disorder only.}
\label{totsfd_comp}
\end{figure}

\par{Figure \ref{pm_ke_comp} (a) shows the local kinetic energy considering low on-site disorder ($\sigma = 0.05$) at the boron sites only while Fig. \ref{pm_ke_comp} (b) shows the corresponding figure considering on-site disorder at all lattice sites (with $\sigma$ = 0.5). The relative uniformity of the kinetic energy specifically in the superconducting regions illustrates the robustness of the local kinetic energy to on-site disorder. However considering only structural disorder with a disorder parameter of 0.5 (Fig. \ref{pm_ke_comp} (c)) we find that there is a large variation in the local kinetic energy, so much so that the energy scale is very different to the case of on-site disorder at all sites. This reflects the more drastic changes in the kinetic energy which accompany changes in the hopping parameter and the concurrent overall increase in the kinetic energy with increasing structural disorder.}

\par{Figure \ref{pm_ke_comp} (d) shows the paramagnetic response considering low on-site disorder ($\sigma = 0.05$) at the boron sites only and illustrates uniformity within the superconducting regions. Considering on-site disorder at all lattice sites (Fig. \ref{pm_ke_comp} (e) with $\sigma = 0.5$) we find a relatively small change in the paramagnetic response  of superconducting regions. However considering structural disorder only (Fig. \ref{pm_ke_comp} (f) with $\sigma = 0.5$) shows a marked difference to on-site disorder. The paramagnetic response clearly shows an overall increase with enhanced connectivity between superconducting regions. }

\par{These inherent differences between on-site and structural disorder are further highlighted considering the total superfluid density as a function of on-site disorder only (Fig. \ref{totsfd_comp} (a)) and structural disorder only (Fig. \ref{totsfd_comp} (b)). While on-site disorder results in a monotonic decrease in the superfluid density, structural disorder shows an initially rapid decrease followed by an increase however the superfluid density is quite small. Initially, the kinetic energy changes little with structural disorder (not shown here) however at higher structural disorder, the kinetic energy begins to increase, as does the paramagnetic response resulting in the initial drop followed by an increase.}

\section{Discussion}

\par{The variation of the structural disorder parameter with boron concentration was motivated by examining Raman studies of BNCD which show a change in the $D$ and $G$ peaks as the boron concentration is changed \cite{Gajewski2009} although the significance of the structural change has not yet been established \cite{Zhang2014}. A greater change in the Raman spectra has been found in nitrogen doped carbon films and nitrogen doped nano-crystalline diamond films \cite{mkrmsb} where nitrogen incorporation significantly modifies the microstructure of these films. The incorporation of boron modifies the diamond lattice in the regions where the boron is substitutionally incorporated. We assume that the disorder parameter may vary linearly, exponentially or inverse exponentially along with the boron concentration as there is yet insufficient Raman analysis available to specify the change in the level of structural disorder with the boron concentration. The three different cases can however be compared to experimental data of the variation of the $T_{c}$ with the boron concentration which is also comparable to the behaviour found theoretically using the Berlitz theory (which is an application of the McMillan formula for disordered systems) \cite{Bretislav2009}.}

\par{The significance of structural disorder can be determined by comparing the distribution of local pairing amplitudes for the different cases of on-site potential disorder and structural disorder. In the case of on-site disorder, there is a decrease in the number of sites with pairing amplitude around 1.6$t$ while the number of sites with a local pairing amplitude of zero increases only slightly. Conversely, in the case of structural disorder the number of sites with local pairing amplitude close to 1.6$t$ decreases rapidly while the number of sites with non-zero pairing amplitude increases dramatically, with the number of sites with zero pairing amplitude decreasing. This complete difference in behaviour highlights the different natures of structural and on-site disorder. On-site disorder results in shifts of the eigenenergies at boron sites resulting in little spread of the pairing amplitude while structural disorder can promote connectivity between regions with non-zero pairing amplitude due to the distribution of eigenenergies it creates. Opposite trends in the behaviour of the average kinetic energy have been noted for structural and on-site disorder \cite{Dey2008} where it was found that the average kinetic energy increases for on-site disorder and decreases for structural disorder.}

\par{On-site potential disorder results in enhanced localization of Cooper pairs. On the other hand, structural disorder results in a change in the coupling between different sites, resulting in a greater number of sites with non-zero pairing amplitude due to enhanced connectivity. While the mean pairing amplitude increases with increasing disorder in the case of structural disorder as well as considering on-site disorder at all sites (Figs. \ref{epsdisALL_comb} and \ref{struc_dis}) the increase in the case of structural disorder is far greater, as is the decrease in the spectral gap with disorder. In addition, the distribution of local pairing amplitudes is very different for on-site and structural disorder even for the same values of disorder parameter. On-site disorder largely preserves the local pairing amplitude while structural disorder rapidly decreases the number of sites with relatively large pairing amplitude. This highlights the fundamental difference in the effects of structural and on-site disorder.}

\par{It is significant that in the case of on-site potential disorder the spectral gap  in the density of states does not close (this has been found in other studies as well \cite{Ghosal2001}). It has been shown that in the half-filled Hubbard Hamiltonian the different changes in symmetry introduced by structural and on-site disorder have different effects on the conductivity \cite{Denteneer2001}. Examination of the density of states shows that on-site disorder does not induce available states close to the gap. As the coupling between lattice sites varies, a significant number of eigenstates become available in competition to the energy gained through gap formation. In the extreme case of high structural and on-site disorder, the gap eventually vanishes resulting in a pseudo-gapped ground state.} 

\par{Considering the temperature dependence, experimental scanning transmission spectrosocpy studies of BNCD have shown different temperature dependences of the local energy gap for BCS-like and non-BCS like regions \cite{AchatzPSS}. In addition, the superconducting gap has been found to vary within nano-diamond grains themselves \cite{Willems2009}. The different temperature dependences of the calculated spectral gap and mean pairing amplitudes for structural as opposed to on-site disorder suggest that in regions where structural disorder is dominant, the temperature dependence will not follow the conventional BCS temperature dependence while in regions where structural disorder is less influential than on-site disorder (or where disorder of both kinds is minimal) the temperature dependence will follow the conventional BCS dependence. While the pairing mechanism does not change in these different regions, structurally disordered systems show a significant deviation from the variation in the temperature dependence of the spectral gap which may explain the deviations from the conventionally expected BCS dependence found in BNCD grains experimentally. This would manifest as variations in the gap within the grains, particularly close to grain boundaries and within grain boundaries as found experimentally \cite{AchatzPSS} due to the increased structural disorder and decreased boron concentration in these regions. The differences in the temperature evolution of structural as opposed to on-site disorder shows further deviation from the conventional BCS theory. This is also reflected in the substantial difference between the changes induced in the superfluid density by on-site and structural disorder.}

\par{The initial decrease followed by an increase in the superfluid density with structural disorder is counter-intuitive although it can be understood as follows. Initially, the increase in structural disorder rapidly decreases the number of lattice sites with relatively high local pairing amplitudes so the mean superfluid density rapidly decreases. With further increase in the structural disorder, the number of sites with non-zero pairing continues to increase with the random variation of the pairing amplitudes resulting in increased isotropy as local superconducting islands begin to spread out. This is borne out by the increase in the range of the pairing amplitude correlation function when considering high structural disorder (inset in Fig. \ref{9}). The local paramagnetic response (Fig. \ref{pm_ke_comp} (f)) shows an overall increase with enhanced connectivity between regions with a high paramagnetic response. The average superfluid density, being sensitive to the uniformity of the pairing amplitude across adjacent regions, therefore begins to increase with increasing structural disorder.}

\section{Conclusion}

\par{The contrasting nature of on-site random potential disorder and structural disorder in the form of non-uniform random tight-binding hopping parameters has been illustrated and the results have been applied to understanding the superconductor-insulator transition in BDD and BNCD. While on-site potential disorder at the boron sites alone decreases the mean pairing amplitude and spectral gap, structural disorder can increase the mean pairing amplitude although the spectral gap still decreases as the disorder parameter increases. This significant difference in behaviour stems from the introduction of states within the gap region in the case of structural disorder, resulting in a wide distribution of mean pairing amplitudes. The calculated temperature dependence also highlights the different natures of structural and on-site disorder as structurally disordered systems show a marked deviation from the temperature dependence expected for a conventional BCS superconductor. This work is of particular interest when applied to BNCD due to the high level of structural disorder inherent in this unconventional superconductor. Through the combination of structural disorder, random on-site potential fluctuations and the Coulomb interaction in the narrow acceptor band, the experimentally found saturation of the $ T_{C} $ in BDD as a function of the boron concentration can be understood. This study illustrates a minimal Hamiltonian which captures some of the features found in experimental studies of BDD and BNCD and lays a foundation for further work where grain boundaries in BNCD (which act as weak  links) can be built into the theory through Josephson junctions to more closely reflect BNCD and be more directly comparable to transport and measurements.}

\section{Acknowledgements}
\par{We thank Milo\v{s} Nesl\'{a}dek, Ping Sheng and Swagata Acharya for useful discussions. R.M. and S.B. thank the CTS, IIT Kharagpur for supporting a research visit. We thank the Mathematical Sciences Cluster, Wits for computational resources and Shunmuga Pillay for support. S.B. would like to thank the National Research Foundation (SA) for granting the Nanotechnology Flagship Programme to perform this work and also the University of the Witwatersrand Research Council for the financial support.}


\begin{thebibliography}{39}%
\bibitem{AMTrans}
D. Belitz and T. R. Kirkpatrick,
\newblock Rev. Mod. Phys. {\bf 66,}, 261 (1994).
\bibitem{AndTrans}
F. Evers and A. D. Mirlin,
\newblock Rev. Mod. Phys. {\bf 80}, 1355 (2008).
\bibitem{GantRev}
V. F. Gantmakher and V. T. Dolgopolov,
\newblock Phys. Usp. {\bf 53}, 1 (2010).
\bibitem{Denteneer2001}
P. J. H. Denteneer, R. T. Scalettar, and N. Trivedi,
\newblock Phys. Rev. Lett. {\bf 87}, 146401 (2001).
\bibitem{Jiang2013}
M. Jiang, R. Nanguneri, N. Trivedi, G. G. Batrouni, and
R. T. Scalettar,
\newblock New J. Phys. {\bf 15}, 023023 (2013).

\bibitem{Trivedi1996R}
N. Trivedi, R. T. Scalettar, and M. Randeria,
\newblock Phys. Rev. B {\bf 54}, R3756 (1996).

\bibitem{SacepePRL2008}
B. Sac\'{e}p\'{e}, C. Chapelier, T. I. Baturina, V. M. Vinokur,
M. R. Baklanov, and M. Sanquer,
\newblock Phys. Rev. Lett. {\bf 101}, 157006 (2008).

\bibitem{Potirniche2014}
I.-D. Potirniche, J. Maciejko, R. Nandkishore, and S. L.
Sondhi,
\newblock Phys. Rev. B {\bf 90}, 094516 (2014).

\bibitem{Mohanta2015}
N. Mohanta and A. Taraphder
\newblock Phys. Rev. B {\bf 92}, 174531 (2015).

\bibitem{Ghosal1998}
A. Ghosal, M. Randeria, and N. Trivedi,
\newblock Phys. Rev. Lett. {\bf 81}, 3940 (1998).

\bibitem{Chatterjee2008582}
B. Chatterjee and A. Taraphder,
\newblock Solid State Commun. {\bf 148}, 582 (2008).

\bibitem{Ghosal2001}
A. Ghosal, M. Randeria, and N. Trivedi,
\newblock Phys. Rev. B {\bf 65}, 014501 (2001).

\bibitem{Cerovski1999}
V. Z. Cerovski, S. D. Mahanti, T. A. Kaplan, and A. Taraphder,
\newblock Phys. Rev. B {\bf 59}, 13977 (1999).

\bibitem{Takano2004}
Y. Takano, M. Nagao, I. Sakaguchi, M. Tachiki, T. Hatano,
K. Kobayashi, H. Umezawa, and H. Kawarada,
\newblock Appl. Phys. Lett. {\bf 85}, 2851 (1999).

\bibitem{PhysRevLett.96.097006}
B. Sac\'{e}p\'{e}, C. Chapelier, C. Marcenat, J. Ka\v{c}mar\v{c}ik,
T. Klein, M. Bernard, and E. Bustarret,
\newblock Phys. Rev. Lett. {\bf 96}, 097006 (2006).

\bibitem{Baskaran}
G. Baskaran,
\newblock J. Supercond. Nov. Magn. {\bf 21}, 45 (2008).

\bibitem{Achatz2010814}
P. Achatz, F. Omns, L. Ortga, C. Marcenat, J. Vack,
V. Hnatowicz, U. Kster, F. Jomard, and E. Bustarret,
\newblock Diamond Relat. Mater. {\bf 19}, 814 (2010).

\bibitem{Dubrovinskaia}
N. Dubrovinskaia, L. Dubrovinsky, T. Papageorgiou,
A. Bosak, M. Krisch, H. F. Braun, and J. Wosnitza,
\newblock Appl. Phys. Lett. {\bf 92}, 132506 (2008).

\bibitem{Ekimov2004}
E. Ekimov, V. Sidorov, E. Bauer, N. Melnik, N. Curro,
J. Thompson, and S. Stishov,
\newblock Nature {\bf 482}, 542 (2004).

\bibitem{PhysRevLett.93.237004}
X. Blase, C. Adessi, and D. Conn\'{e}table
\newblock Phys. Rev. Lett. {\bf 93}, 237004 (2004).



\bibitem{Kawano2010}
A. Kawano, H. Ishiwata, S. Iriyama, R. Okada, T. Yam-
aguchi, Y. Takano, and H. Kawarada,
\newblock Phys. Rev. B {\bf 82}, 085318 (2010).

\bibitem{Shirakawa}
T. Shirakawa, S. Horiuchi, Y. Ohta, and H. Fukuyama,
\newblock Phys. Soc. Jpn {\bf 76}, 014711 (2007).

\bibitem{Okazaki2011582}
H. Okazaki, T. Arakane, K. Sugawara, T. Sato, T. Taka-
hashi, T. Wakita, M. Hirai, Y. Muraoka, Y. Takano,
S. Ishii, S. Iriyama, H. Kawarada, and T. Yokoya,
\newblock J. Phys. Chem. Solids {\bf 72}, 582 (2011).

\bibitem{AchatzPSS}
P. Achatz, E. Bustarret, C. Marcenat, R. Piquerel,
T. Dubouchet, C. Chapelier, A. M. Bonnot, O. A.
Williams, K. Haenen, W. Gajewski, J. A. Garrido, and
M. Stutzmann,
\newblock phys. status solidi (a) {\bf 206}, 1978 (2009).

\bibitem{PhysRevB.82.033306}
F. Dahlem, P. Achatz, O. A. Williams, D. Araujo, E. Bus-
tarret, and H. Courtois,
\newblock Phys. Rev. B {\bf 82}, 033306 (2010).

\bibitem{Ohta2007121}
Y. Ohta, T. Shirakawa, S. Horiuchi, and H. Fukuyama,
\newblock Physica C: Supercond. {\bf 460462}, 121 (2007).

\bibitem{Klein2007}
T. Klein, P. Achatz, J. Kacmarcik, C. Marcenat,
F. Gustafsson, J. Marcus, E. Bustarret, J. Pernot,
F. Omnes, B. E. Sernelius, C. Persson, A. Ferreira da Silva,
and C. Cytermann,
\newblock Phys. Rev. B {\bf 75}, 165313 (2007).

\bibitem{BustarretPRL}
E. Bustarret, J. Ka\v{c}mar\v{c}ik, C. Marcenat, E. Gheeraert,
C. Cytermann, J. Marcus, and T. Klein,
\newblock Phys. Rev. Lett. {\bf 93}, 237005 (2004).

\bibitem{Winzer200565}
K. Winzer, D. Bogdanov, and C. Wild,
\newblock Physica C {\bf 432}, 65 (2005).
\bibitem{Bretislav2009}
B. \v{S}opik,
\newblock New J. Phys. {\bf 11}, 103026 (2009).
\bibitem{PhysRevB.77.064518}
J. E. Moussa and M. L. Cohen,
\newblock New J. Phys. {\bf 77}, 064518 (2008).
\bibitem{Das}
T. Das, J.-X. Zhu, and M. J. Graf,
\newblock Phys. Rev. B {\bf 84}, 134510 (2011).

\bibitem{Huang}
H. Huang, Y. Gao, J.-X. Zhu and C. S. Ting, C. S.,
\newblock Phys. Rev. Lett. {\bf 109}, 187007 (2012).

\bibitem{BC}
B. Chattopadhyay, D. M. Gaitonde and A. Taraphder, 
\newblock Europhys. Lett. {\bf 34}, 705 (1996). 

\bibitem{mkrmsb}
M. V. Katkov, R. McIntosh, and S. Bhattacharyya,
\newblock J. Appl. Phys. {\bf 113}, 093701 (2013).

\bibitem{Alibart2010}
F. Alibart, M. Lejeune, O. Durand Drouhin, K. Zellama,
and M. Benlahsen,
\newblock J. Appl. Phys. {\bf 108}, 053504 (2010).


\bibitem{Shah2010}
K. V. Shah, D. Churochkin, Z. Chiguvare, and S. Bhat-
tacharyya,
\newblock J. Appl. Phys. {\bf 82}, 184206 (2010).

\bibitem{Bhat2008}
S. Bhattacharyya,
\newblock Phys. Rev. B {\bf 77}, 233407 (2008).

\bibitem{TrivediNatphys}
K. Bouadim, Y. L. Loh, M. Randeria and N. Trivedi
\newblock Nat. Phys. {\bf 7}, 884 (2011).

\bibitem{Gajewski2009}
W. Gajewski, P. Achatz, O. A. Williams, K. Haenen,
E. Bustarret, M. Stutzmann, and J. A. Garrido,
\newblock Phys. Rev. B {\bf 79}, 045206 (2009).

\bibitem{Zhang2014}
G. Zhang, S. Turner, E. A. Ekimov, J. Vanacken, M. Tim-
mermans, T. Samuely, V. A. Sidorov, S. M. Stishov, Y. Lu,
B. Deloof, B. Goderis, G. Van Tendeloo, J. Van de Vondel,
and V. V. Moshchalkov,
\newblock Adv. Mater. {\bf 26}, 2034 (2014).

\bibitem{Dey2008}
P. Dey and S. Basu,
\newblock J. Phys. Condens. Matter {\bf 20}, 485205 (2008).

\bibitem{Willems2009}
B. L. Willems, V. H. Dao, J. Vanacken, L. F. Chibotaru,
V. V. Moshchalkov, I. Guillam\'{o}n, H. Suderow, S. Vieira,
S. D. Janssens, O. A. Williams, K. Haenen, and P. Wag-
ner,
\newblock Phys. Rev. B {\bf 80}, 224518 (2009).





\end{thebibliography}


%


\end{document}